\PassOptionsToPackage{unicode}{hyperref}
\PassOptionsToPackage{hyphens}{url}
\documentclass[
]{article}
\usepackage{xcolor}
\usepackage[margin=1in]{geometry}
\usepackage{amsmath,amssymb}
\usepackage{iftex}
\ifPDFTeX
  \usepackage[T1]{fontenc}
  \usepackage[utf8]{inputenc}
  \usepackage{textcomp} 
\else 
  \usepackage{unicode-math} 
  \defaultfontfeatures{Scale=MatchLowercase}
  \defaultfontfeatures[\rmfamily]{Ligatures=TeX,Scale=1}
\fi
\usepackage{lmodern}
\ifPDFTeX\else
\fi
\IfFileExists{upquote.sty}{\usepackage{upquote}}{}
\IfFileExists{microtype.sty}{
  \usepackage[]{microtype}
  \UseMicrotypeSet[protrusion]{basicmath} 
}{}
\makeatletter
\@ifundefined{KOMAClassName}{
  \IfFileExists{parskip.sty}{%
    \usepackage{parskip}
  }{
    \setlength{\parindent}{0pt}
    \setlength{\parskip}{6pt plus 2pt minus 1pt}}
}{
  \KOMAoptions{parskip=half}}
\makeatother
\usepackage{graphicx}
\makeatletter
\newsavebox\pandoc@box
\newcommand*\pandocbounded[1]{
  \sbox\pandoc@box{#1}%
  \Gscale@div\@tempa{\textheight}{\dimexpr\ht\pandoc@box+\dp\pandoc@box\relax}%
  \Gscale@div\@tempb{\linewidth}{\wd\pandoc@box}%
  \ifdim\@tempb\p@<\@tempa\p@\let\@tempa\@tempb\fi
  \ifdim\@tempa\p@<\p@\scalebox{\@tempa}{\usebox\pandoc@box}%
  \else\usebox{\pandoc@box}%
  \fi%
}
\def\fps@figure{htbp}
\makeatother
\usepackage[numbers]{natbib}
\usepackage{caption} 

\usepackage{makecell}

\providecommand{\bm}[1]{\symbf{#1}}
\renewcommand{\bm}[1]{\symbf{#1}}

\usepackage{booktabs}
\usepackage{longtable}
\usepackage{array}
\usepackage{multirow}
\usepackage{wrapfig}
\usepackage{float}
\usepackage{colortbl}
\usepackage{pdflscape}
\usepackage{tabu}
\usepackage{threeparttable}
\usepackage{threeparttablex}
\usepackage[normalem]{ulem}
\usepackage{makecell}
\usepackage{xcolor}
\usepackage{bookmark}
\IfFileExists{xurl.sty}{\usepackage{xurl}}{} 
\hypersetup{
  pdfauthor={Mohsen Sadatsafavi, Richard D. Riley},
  pdfkeywords={Bayesian inference; Clinical prediction models;
Uncertainty quantification; Prognosis; Predictive value of tests},
  hidelinks,
  pdfcreator={LaTeX via pandoc}}

\title{\centering
\parbox{\linewidth}{
  \centering
  Progression to the mean:\\[6pt] 
  A comparison of Bayesian clinical prediction models outputting the posterior mean versus conventional plug-in predictions
}}
\author{Mohsen Sadatsafavi, Richard D. Riley}
\date{2026-06-08}

\begin{document}
\maketitle
\begin{abstract}
Clinical prediction models provide predictions for individuals,
typically expressed as point estimates derived from a deterministic
function, such as a logistic regression equation. Such `plug-in'
predictions hide inherent uncertainty. In contrast, Bayesian methods
offer a coherent mechanism for uncertainty propagation, and allow the
computation of the posterior mean as the measure of centrality of choice
for clinical decision-making. However, Bayesian methods are not widely
utilised in predictive analytics for healthcare. We investigated the
feasibility and performance of a Bayesian adaptation of the commonly
used frequentist framework for risk prediction modelling. We assessed
(i) the use of shrinkage priors with complementary features (simplicity,
user input, and automatic shrinkage) that enable Laplace/normal
approximation of the posterior, and (ii) exact and approximate methods
for efficient computation of the posterior mean. Using examples and
simulations, we demonstrate that this Bayesian approach is feasible and
improves predictive performance, while enabling uncertainty
quantification with suitable coverage. In small-to-medium sample sizes,
the gain in clinical utility by using the posterior mean over plug-in
predictions was equivalent to the gain from using a noticeably larger
sample size. Adapting the widely used parametric regression methods to
an approximate Bayesian framework for prediction modelling is both
pragmatic and clinically advantageous.
\end{abstract}

\let\thefootnote\relax\footnotetext{From Respiratory Evaluation Sciences Program, Faculty of Medicine and Faculty of Pharmaceutical Sciences, The University of British Columbia (MS); Department of Applied Health Sciences, School of Health Sciences, College of Medicine and Health, University of Birmingham (RDR)}

\let\thefootnote\relax\footnotetext{* Correspondence to Mohsen Sadatsafavi, Room 4110, 2405 Wesbrook Mall, Vancouver, British Columbia, V6T1Z3, Canada; email: mohsen.sadatsafavi@ubc.ca}

\clearpage

\section{Background}\label{background}

The prevailing practice in predictive analytics for healthcare is to
derive and present a prediction model as a deterministic function,
resulting in a single predicted value for an individual. An example is a
prediction equation corresponding to a fitted logistic regression model
that provides a single point estimate of an individual's outcome
probability (risk) conditional on a function of multiple predictors.
These point estimates are often based on the maximum likelihood
estimates (MLE) of regression coefficients, perhaps after applying some
form of shrinkage. Focusing solely on such point estimates obscures the
reality that model-based predictions are inherently uncertain, as models
are developed using finite
samples\citep{riley_importance_2025, riley_multiverse_2023}.

The relevance of reporting uncertainty around predicted values at the
time of deploying a clinical prediction model is
recognised\citep{thomassen_effectiven_2024, riley_uncertainty_2025}. It
is argued that model developers have an ethical responsibility towards
quantifying the strength of evidence behind a model's predictions,
including uncertainty quantification at the point of care, which can in
turn help with critical appraisal of a model, contribute towards
assessments of model fairness, and facilitate shared
decision-making\citep{riley_uncertainty_2025}. Reporting and
communicating prediction uncertainty, however, requires changes in
current practices. For example, when the model is based on multivariable
regression, the uncertainty around parameter estimates (e.g., the
covariance matrix) should be deployed alongside the prediction equation,
allowing the quantification of prediction uncertainty for an individual.

Such a change in practice opens a unique window of opportunity for
adopting Bayesian thinking in prediction modelling. Bayesian methods not
only facilitate intuitive representation of uncertainty, but also enable
more principled decision-making, in particular as they enable the
computation of the posterior mean. Exact Bayesian inference often
requires Markov Chain Monte Carlo (MCMC)-based sampling approaches.
These techniques can be seen as the gold standard for Bayesian inference
as they enable draws from the exact posterior distributions. They also
enable flexible modelling by admitting a vast class of priors and link
functions. This frees the investigator to construct potentially complex
models that best approximate the reality, and couple them with
sparsity-inducing priors. From a pragmatic perspective, however, a model
in this scheme is in effect a large set of posterior draws, which might
complicate presentation, deployment, and transportability across
populations in a reproducible manner. This may be preventing healthcare
researchers from using Bayesian regression methods for developing
clinical prediction models; indeed, most applied investigators seem to
favour frequentist regression modelling, perhaps due to the objectivity,
user-friendliness, reproducibility, transparency, and usability of the
current MLE-based machinery in prediction modelling.

To address this, this article examines the performance of an approximate
Bayesian approach to model development that maintains these features of
current MLE-based clinical prediction models, whilst also allowing
coherent uncertainty propagation and the output of the posterior mean.
We do not claim novelty in the application of Bayesian methods in
clinical
prediction\citep{kazemi_improving_2012, fernandes_risk_2017, debray_bayescpm_2012, murphy_bayesian_2019, zhao_bayesian_2023, hageman_estimating_2024, vialard_developing_2025},
nor do we intend to rehash the extensive literature on Bayesian
predictive inference\citep{gelman_bayesian_2014}. Instead, we
investigate the feasibility and performance of a general framework for
developing risk prediction models that encompasses the main principles
of Bayesianism while remaining compatible with familiar risk modelling
tools, making it more immediately applicable to statisticians, data
scientists, and other healthcare researchers developing and deploying
prediction models.

The outline of this paper is as follows. We briefly review key
frequentist and Bayesian approaches for predictive inference. We recap
the central role of posterior mean of outcome probability from an
expected utility perspective. We then discuss how conventional
regression-based approaches for risk prediction modelling can be tweaked
to enable Bayesian inference. We discuss the two key components:
specifying priors that enable Laplace approximation of the posterior,
and various approaches for deployment-time computation of the posterior
mean. We run simulation studies that show this approach offers
performance that can exceed that of plug-in predictions from commonly
used frequentist methods while enabling coherent uncertainty
propagation, and seems to be comparable with fully Bayesian MCMC-based
approaches while remaining pragmatic. We showcase the developments in an
illustrative example, and conclude by highlighting some open questions
and areas for future research.

\section{An overview of frequentist and Bayesian frameworks for
developing a clinical prediction
model}\label{an-overview-of-frequentist-and-bayesian-frameworks-for-developing-a-clinical-prediction-model}

In this section, we outline the proposed pipeline for modelling the
probability (risk) of a binary outcome, but the methodology is general
and extendable to other outcome types. Consider a prediction model
\(\pi(x;\bm{\beta})\), a deterministic function that estimates the
probability of a clinically important outcome given an individual's
characteristics (predictors) \(\mathbf x\), indexed by parameters
\(\bm{\beta}\). This model is produced using a development sample \(D\)
containing predictor and outcome information for \(n\) individuals. By
\(\pi_i\) we refer to a predicted value of outcome probability for the
\(i\)th individual. Our focus is on parametric regression models, where
\(\bm{\beta}\) maps to regression coefficients. For example, for a
logistic regression model, \(\bm{\beta}\) will include the intercept
term and log-odds-ratios for predictors, and this leads to a predicted
outcome probability of
\(\pi(x;\bm{\beta})=(1+e^{-x^T \bm{\beta}} )^{-1}\). For proportional
hazards models, it includes the (log) cumulative baseline hazard at the
time point of interest and log-hazard-ratios for covariates.

\subsection{The frequentist framework and plug-in
predictions}\label{sec-freq}

In conventional practice, prediction models are presented using a single
set of coefficients \(\mathbf{\bm{\hat{\beta}}}\). Because most
parametric model specifications allow for the convenient estimation of a
smooth likelihood \(p(D|\bm{\beta})\) and its gradient, deterministic
gradient ascent methods are used to find the MLE of the model parameters
(e.g., intercept and predictor effects):

\[\mathbf{\bm{\hat{\beta}}} = \arg \max p(D|\bm{\beta}).\]

By reporting only \(\mathbf{\bm{\hat{\beta}}}\), we collapse the
information contained in the likelihood function into a single point
estimate for each parameter, which leads to the deterministic function
that makes point estimate predictions for individuals. We refer to these
predictions as plug-in estimates. Uncertainty around
\(\bm{\hat{\beta}}\) is often quantified as the covariance matrix
\(\bm{\Sigma}\) of regression
coefficients\citep{riley_uncertainty_2025}. \(\bm{\Sigma}\) captures the
curvature of the likelihood around the MLE. In the classical normal
approximation of the likelihood function underlying Wald-type
inference\citep{wald_tests_1943}, \(\bm{\beta}\) is assumed to follow
asymptotically a multivariate normal distribution:

\[\bm{\beta} \sim N(\mathbf{\bm{\hat{\beta}}}, \bm{\Sigma}).\]

This approximation is reliable under regularity conditions, e.g.,
\(n \gg\) number of predictors, non-separability, and lack of perfect
collinearity - conditions typically satisfied in well-considered model
development situations with sufficient sample size.

For Generalised Linear Models (GLMs), the multivariate-normality of
\(\bm{\Sigma}\) implies the linear predictor (i.e., plug-in predicted
value for an individual based on the deterministic function on its
original scale, e.g., logit) \(\eta\) follows a univariate normal
distribution:

\[\eta \sim N(\tilde{\bm{x}}^T \mathbf{\bm{\hat{\beta}}}, \tilde{\bm{x}}^T \bm{\Sigma} \tilde{\bm{x}}),\]

with \(\tilde{\mathbf{x}}\) being the vector of predictors for a new
individual. The variance of individual-level predictions on the
transformed scale of interest (probability) can be computed, for
example, using the delta method. More commonly, confidence bands for
predicted probabilities are computed from the corresponding tail
probabilities of \(\eta\), which are preserved under monotonic link
functions.

If \(\mathbf{\bm{\hat{\beta}}}\) is obtained from unpenalised MLE, the
model is at risk of overfitting: predictions tend to be extreme, with
probabilities for high-risk individuals overestimated and those for
low-risk individuals underestimated. Because of such overfitting
concerns, some form of penalisation of the likelihood is often applied.
These penalisations introduce an intentional shrinkage bias in the
choice of \(\bm{\hat{\beta}}\), which tends to reduce overfitting.
Simple, tuning-free penalisations include the Firth correction
(discussed below). More commonly, hierarchical shrinkage methods such as
ridge or LASSO are considered, where a global tuning parameter
determines the strength of penalisation. This parameter is commonly
estimated from the same development data via cross-validation. This
makes standard covariance-based inference on the coefficients invalid,
because the tuning parameter is considered fixed when it is itself a
function of the data. Commonly, in prediction models based on ridge or
LASSO, uncertainty is characterised via bootstrapping. This is
computationally intensive, introduces Monte Carlo noise, and might
result in convergence issues with smaller datasets. For LASSO, some
coefficients might shrink to zero in some iterations, which causes
challenges as to how to summarise the zero-inflated empirical
distribution of \(\bm{\beta}\)s. Recent developments have involved
specialised post-shrinkage inference
techniques\citep{taylor_statistical_2015}. But these approaches target
coverage-correct confidence intervals on regression coefficients, not on
individual-level predicted probabilities, which are what clinical
decision-making requires.

\subsection{The Bayesian framework and predictions based on the
posterior
mean}\label{the-bayesian-framework-and-predictions-based-on-the-posterior-mean}

In the Bayesian approach, regression coefficients are treated as random
variables, and the posterior distribution given the data is obtained via
Bayes' rule: we place a prior on \(\bm{\beta}\), \(p(\bm{\beta})\), that
is combined with the likelihood to estimate the posterior probability
\(p(\bm{\beta}|D)\):

\[p(\bm{\beta}|D) \propto p(\bm{\beta})p(D|\bm{\beta}).\]

Therefore, in a regression framework, one extra step in the Bayesian
thinking is the specification of a prior on regression coefficients such
as intercepts and predictor effects.

Indeed, placing an improper flat prior (\(p(\bm{\beta}) \propto 1\)) on
coefficients provides a direct Bayesian interpretation of the
frequentist estimation of the likelihood. In this case, the posterior is
proportional to the likelihood, and the multivariate normal
approximation of the likelihood can be seen as a Laplace approximation
of the posterior distribution\citep[section 4.1]{gelman_bayesian_2014}.
As such, the MLE estimates of regression coefficients become maximum a
posteriori (MAP) estimates under the Bayesian framework, and Wald-type
CIs can be interpreted as approximate CrIs.

For the more general case, if the prior and the likelihood are
conjugate, the posterior distribution can be expressed in closed form
and inference can be analytically tractable. In many instances,
conjugacy is not available. General-purpose Bayesian inference platforms
that offer flexibility in the choice of the prior often perform
inference using MCMC methods, resulting in draws from
\(p(\bm{\beta}|D)\). Alternatively, one can place an improper prior on
the data-generating mechanism itself, giving rise to the Bayesian
bootstrap\citep{rubin_bayesian_1981}. In this approach, we obtain a
posterior distribution over the population from which the sample was
drawn, which then propagates to the parameters (e.g., regression
coefficients).

From a given posterior distribution of regression coefficients, two
distinct predictions can be derived. The plug-in estimate,
\(\pi(\mathbf{x}; \mathbf{\bm{\hat{\beta}}})\), uses only the MAP value
for each parameter and ignores uncertainty; this coincides conceptually
with the prediction one would obtain in a frequentist framework (see
section \ref{sec-freq}). We note that under a Laplace approximation,
\(\bm{\hat{\beta}}\) is the posterior mean, median, and mode of
\(p(\bm{\beta}|D)\). However, due to the monotonic but non-linear link
function, the predicted probabilities using this plug-in approach only
remain the posterior median of their respective posterior distribution,
but not its mean or mode. On the other hand, the posterior mean,
\(\mathbb{E}[\pi(\mathbf{x};\bm{\beta})]\), integrates over the full
posterior of \(\bm{\beta}\), incorporating parameter uncertainty into
the prediction. For the remainder of this paper, we refer to these two
types of predictions as plug-in point estimates (PE), based on the
deterministic function of \(\bm{\hat{\beta}}\), and posterior mean (PM),
taken as the mean from the individual's posterior distribution of
predicted probabilities. Under standard regularity conditions, different
summaries of the posterior distribution (e.g., mean, median, and mode)
converge to the same estimand asymptotically. However, their
finite-sample behaviour can differ.

\subsection{Posterior mean as the central quantity for clinical
decision-making}\label{posterior-mean-as-the-central-quantity-for-clinical-decision-making}

In Bayesian inference, the most common measure of centrality is the
posterior mean:

\[\mathbb{E}(\pi|\tilde{\mathbf{x}}) = \int \pi(\tilde{\mathbf{x}}; \bm{\beta}) p(\bm{\beta}|D) d\bm{\beta}.\]

Bayesian decision theory provides a formal justification for this
quantity as the optimal measure of centrality. It is the unique
minimiser of several important classes of loss functions, namely the
quadratic and the information-theoretic
loss\citep{berger_statistical_2006}. Of particular importance are
information-theoretic loss functions, as they are deeply connected to
statistical inference. Likelihood-based methods, whether frequentist or
Bayesian, can be interpreted through the lens of minimising information
loss between the candidate model and the true data-generating
process\citep{berger_statistical_2006}. Such connections to
information-theoretic concepts become relevant in this paper when we
approximate the posterior mean and investigate the coverage of the
posterior distribution.

In the context of clinical decision-making, the posterior mean can be
further defended as the singular metric of importance within an expected
utility framework. While different versions of this argument appear in
several classical
papers\citep{pauker_therapeutic_1975, vickers_decision_2006, metz_roc_1978, phelps_decision_1988},
it might be worth re-expressing this reasoning in the context of
uncertain predicted risks. We follow the `threshold' argument by Pauker
and Kassirer\citep{pauker_therapeutic_1975}, which directly relates to
the concept of net benefit (NB) by Vickers and
Elkin\citep{vickers_decision_2006}. Let \(u_{ab}\) indicate the utility
of offering treatment \(a\) (0: no treatment, 1: treatment) to someone
with future outcome status \(b\) (0: will not experience, 1: will
experience the outcome), with \(u_{11} > u_{01}\) and
\(u_{00} > u_{10}\). The incremental utility of treatment versus no
treatment for someone with outcome probability \(\pi\) can be written
as:

\[\Delta u(\pi) = (u_{11} - u_{01})\pi + (u_{10} - u_{00})(1 - \pi).\]

This can be rearranged into a NB function:

\[NB(\pi) = \pi - \frac{u_{00} - u_{10}}{u_{11} - u_{01}}(1 - \pi).\]

where the ratio \(\frac{u_{00}-u_{10}}{u_{11}-u_{01}}\) represents the
relative harm from unnecessary treatment compared to the benefit of
appropriate treatment. As such, if we define
\(z = \frac{u_{00}-u_{10}}{u_{00}-u_{10}+u_{11}-u_{01}}\), we will have

\[NB(\pi) = \pi - \frac{z}{1-z}(1 - \pi) = \frac{\pi - z}{1 - z}.\]

For a decision-maker maximising the expected utility, the optimal rule
is to treat when the expected NB is positive. As NB is a linear function
of outcome probability, the decision rule will be to treat if

\[\mathbb{E}(\pi) \geq z.\]

Thus, the posterior mean of \(\pi\) is the metric that is needed for
optimal decision-making. As this quantity can differ from the point
estimate from (penalised) MLE, an individual's prediction might change
relative to the treatment threshold. This can be seen for our patient
Sam in Table \ref{tab:gusto_coefs}, across the various model development
approaches (another example is provided in Riley et
al\citep{riley_uncertainty_2025}).

\section{An approximate Bayesian approach for posterior mean predictions
and uncertainty quantification}\label{sec-framework}

Building on the foundations detailed in the previous sections, we now
advocate for a pragmatic Bayesian approach that emphasises objectivity,
reproducibility, and ease of implementation. This approach consists of
two core components. First, it utilises priors that enable Laplace
approximation of the posterior distribution without the need for Monte
Carlo approaches (MCMC sampling or bootstrapping). Second, it encourages
the use of posterior mean as the measure of centrality for informing
individual decision-making, and offers a suite of tools ranging from
precise to approximate methods for its computation.

\subsection{`Practical' priors that enable Laplace approximation of the
posterior
distribution}\label{practical-priors-that-enable-laplace-approximation-of-the-posterior-distribution}

In line with our pragmatic pipeline, we restrict attention to priors
that enable a Laplace / normal approximation of the posterior. This
enables packaging a model for deployment as a vector of point estimate
and a covariance matrix that specify a multivariate normal distribution
for regression coefficients. Without claiming exhaustiveness, we
consider four priors that span a range of shrinkage strategies: an
improper flat prior (no shrinkage), the Jeffreys prior (fixed,
tuning-free shrinkage), the log-F prior (user-specified shrinkage), and
a hierarchical Gaussian prior (Bayesian ridge; data-driven shrinkage).

\subsubsection{Flat prior and the inherent shrinkage properties of the
posterior
mean}\label{flat-prior-and-the-inherent-shrinkage-properties-of-the-posterior-mean}

An improper flat prior on coefficients enables a direct interpretation
of the unpenalised regression in a Bayesian way, providing a convenient
Bayesian interpretation of unpenalised regression. While the flat prior
does not introduce any shrinkage, predictions based on the posterior
mean would often show some shrinkage behaviour compared to using the
plug-in values of regression coefficients. Here we apply the word
shrinkage loosely to invoke the notion that the posterior mean generally
results in less variation in predicted probabilities. This arises from
the fact that the posterior mean is the average of predictions across
the `model space', defined by \(p(\bm{\beta}|D)\). Optimistic
predictions are moderated by contributions from regions of the parameter
space with a poorer fit to the data. For the logit link function, this
can be more formally approached as an application of Jensen's
inequality: the non-linear link function means
\(\mathbb{E}[\pi(\mathbf{x}, \bm{\beta})] \geq \pi(\mathbf{x}, \mathbb{E}[\bm{\beta}])\)
for an individual with \(\eta < 0\), and
\(\mathbb{E}[\pi(\mathbf{x}, \bm{\beta})] \leq \pi(\mathbf{x}, \mathbb{E}[\bm{\beta}])\)
for an individual with \(\eta > 0\). Thus, compared to the plug-in
prediction, the posterior mean is shifted towards \(p=0.5\)
(\(\eta = 0\)), with more extreme values shifted more aggressively (a
visualisation of such shrinkage is offered in the Illustrative Example
below).

The plug-in prediction, due to the score equation for the intercept in
unpenalised logistic regression, guarantees that O/E=1 in the sample.
The inherent shrinkage property of posterior mean means this exact
equality does not hold for the posterior mean. On the other hand, the
posterior mean generally improves the calibration slope (bringing it
closer to 1) compared with plug-in predictions. In some sense, the
posterior mean trades exact O/E preservation for improved
individual-level prediction. These will be systematically investigated
in our simulation studies in Section \ref{sec-simulations}.

\subsubsection{Jeffreys prior}\label{jeffreys-prior}

A popular tuning-free prior with mild shrinkage properties is the
Jeffreys prior, defined as the square root of the Fisher information
matrix \(I(\bm{\beta})\):

\[P(\bm{\beta}) \propto |I(\bm{\beta})|^{\frac{1}{2}}\]

This prior was proposed by Firth as a means of reducing the small-sample
bias of the MLE\citep{firth_bias_1993}. It has been shown that it
improves the performance of logistic regression in small samples and
with complete
separation\citep{kass_selection_1996, kosmidis_bias_2009, vanSmeden_firth_2016}.
However, for binary outcomes, it is a proper Bayesian prior with mild
shrinkage properties, with an appealing feature of being invariant under
a change of the scale of predictors\citep{jeffreys_invariant_1946}.
Generally, it can be applied to any parametric model for which the
Fisher information matrix can be computed. From a practical standpoint,
this prior is generally available in standard statistical software
(e.g., SAS, Stata, and R all have implementations of the logistic
regression with Firth correction).

Jeffreys prior is a global shrinkage prior applied to all coefficients,
including the intercept. As such, it shrinks the posterior mean towards
0.5. This is a feature of this prior, and is expected, compared with the
flat prior, to move O/E further away from 1 but the calibration slope
closer to 1.

\subsubsection{Log-F prior}\label{log-f-prior}

Greenland and Mansournia proposed the log-F distribution as a prior for
logistic regression coefficients with sparse
data\citep{greenland_logF_2015}. The symmetric log-F(\(m, m\)) density
on a log-odds coefficient for \(\beta_j\) is

\[p(\beta_j) \propto \frac{e^{m\beta_j/2}}{(1 + e^{\beta_j})^m}.\]

Larger \(m\) gives stronger shrinkage toward zero: log-F(1,1) is weakly
informative (95\% prior interval for the odds ratio for each predictor
of 1/648 to 648). The investigator can optionally choose stronger
shrinkage (higher \(m\)). The corresponding intervals for log-F(2,2) and
log-F(5,5) are, respectively (1/39, 39), and (1/7.1, 7.1). Unlike the
Jeffreys prior, the log-F prior can be selectively applied to individual
predictors, sparing the intercept to avoid shrinking the posterior mean
to 0.5. One can use this prior to incorporate knowledge from domain
experts; for example, to place plausibility bounds on predictor effects.
Log-F priors satisfy the pragmatism of our proposal as they are easy to
implement via data augmentation: a symmetric log-F(\(m, m\)) prior on
\(\bm{\beta}_j\) is obtained by adding two pseudo-observations with
\(x_j=1\), zeros in all other covariate columns (including the
intercept), \(y=0\) in one, and \(y=1\) in the other, and weights
\(m/2\). This makes the log-F family available in any logistic
regression software that accepts observation weights.

\subsubsection{Gaussian prior (approximate Bayesian
ridge)}\label{gaussian-prior-approximate-bayesian-ridge}

In this approach, each coefficient is assigned a zero-mean Gaussian
prior,

\[p(\beta_j) \propto e^{-\frac{\lambda \beta_j^2}{2}},\]

and the intercept is given a flat prior. The precision \(\lambda\)
controls the strength of shrinkage, unlike the log-F prior, where the
shrinkage strength is specified by the user. This makes this prior a
hierarchical one, with the degree of shrinkage decided by the data. A
naïve approach would compute the posterior over \(\bm{\beta}\)
conditional on a single value of \(\lambda\) that maximises the
(marginal) likelihood. This is indeed the same problem that invalidates
the ordinary Laplace-based inference for the frequentist ridge and
LASSO. In our Bayesian setup, the uncertainty in \(\lambda\) must be
propagated into the posterior covariance of \(\bm{\beta}\). Wood
proposed using a Laplace approximation that incorporates the
second-order curvature of the marginal likelihood with respect to
\(\log(\lambda)\), thus adding a correction term to the covariance of
\(\bm{\beta}\)\citep{wood_mgcv_2016}. This yields an estimator that
approximates the posterior without requiring MCMC. This approach is
implemented in the \(\texttt{mgcv}\) R package (and is invoked when
unconditional covariance matrices or predictions are requested). The
same approach is not directly extensible to LASSO, given its posterior
can have point mass at zero for shrunk coefficients, which invalidates
the regularity conditions for Laplace approximation. The Bayesian ridge
is also more objective than the frequentist ridge, with no
cross-validation noise, or need to choose among tuning criteria (e.g.,
deviance, c-statistic, or prediction error). As is standard in penalised
regression, we suggest applying a single global \(\lambda\) to all
coefficients on the standardised scale.

\subsection{Deployment calculations of the posterior
mean}\label{deployment-calculations-of-the-posterior-mean}

Regardless of which of the aforementioned priors is chosen, the Laplace
approximation means the final model can be packaged as a vector of point
parameter estimates and a covariance matrix specifying a multivariate
normal distribution. The multivariate normality of \(p(\bm{\beta}|D)\)
translates to the univariate normality of the linear terms \(\eta\).
Computing the posterior mean for a given patient thus requires computing
the expected value of a transformed normal distribution. For the
logistic regression, this becomes the expectation of a logit-normal
variable, which does not have an exact closed
form\citep{holmes_moments_2022}. Below we provide approaches that strike
different trade-offs between deployment complexity and precision.

\subsubsection{Quadrature}\label{quadrature}

Gauss-Hermite quadrature converts the integrals of type
\(\int_{-\infty}^{\infty} e^{-x^2} f(x)dx\) to weighted sums of \(f()\)s
at pre-defined places on \(x\). For logistic regression, given
\(\eta \sim \text{Normal}(\mu, \sigma)\), this becomes (\(\pi\) is the
mathematical constant, not to be confused with predicted probabilities):

\[\mathbb{E}[\mbox{logit}^{-1}(\eta)] \approx \frac{1}{\sqrt{\pi}} \sum_{k=1}^K w_k \mbox{logit}^{-1}(\mu + \sqrt{2} \sigma x_k),\]

where \(w_k\) and \(x_k\) are the fixed weight and location values. With
\(K\) of 20-30, this approach is effectively indistinguishable from
high-precision numerical integration\citep{liu_gausshermite_1994}. In
this approach, pre-specified location/weight parameters are shipped
alongside the model (2K scalars, which are fixed and can be shared
across models). If uncertainty intervals are to be communicated,
\(\sigma\) is already computed, and so computing the posterior mean will
require a computationally trivial weighted sum of inverse-logit
evaluations.

\subsubsection{MacKay's approximation}\label{mackays-approximation}

An alternative to the quadrature method is the MacKay formula, which
approximates the expectation with one inverse-logit
evaluation\citep{mackay_evidence_1992} (again, \(\pi\) is the
mathematical constant):

\[\mathbb{E}[\mbox{logit}^{-1}(\eta)] \approx \mbox{logit}^{-1} \left( \mu \left( 1 + \frac{\pi \sigma^2}{8} \right)^{-\frac{1}{2}} \right).\]
With \(\sigma\) already computed, this becomes a simple inverse-logit
evaluation.

\subsubsection{Self-projection of mean predicted
probabilities}\label{sec-projpred}

If we plan to deploy a simple, explainable equation that relates
predictors to the expected outcome probability, we can borrow from the
projection predictive literature\citep{piironen_comparison_2017}. As a
brief overview, projection predictive inference attempts to minimise the
discrepancy between a reference and a candidate model. This discrepancy
is measured in terms of KL divergence (\(D_{KL}\)), which can be
interpreted as the extra units of information (bits in base 2 or nats in
base \(e\)) needed to encode a random variable from the reference
distribution using a candidate distribution. For binary responses, if
predicted probabilities for the reference and candidate models are,
respectively, \(p\) and \(q\):

\[D_{KL}(p \parallel q) = p \left[ \log \left( \frac{p}{q} \right) \right] + (1 - p) \left[ \log \left( \frac{1 - p}{1 - q} \right) \right].\]

The overall divergence is the expectation of this quantity over the case
mix (\(p(\bm{x})\)), approximated by the case-mix of the development
sample. In our context, \(p\) and \(q\) are, respectively, the posterior
means, denoted by \(\bar{\pi}_i\), and predictions from the new
approximating function, denoted by \(h(\bm{x}_i; \dot{\bm{\beta}})\)
indexed by parameters \(\dot{\bm{\beta}}\). After removing the terms
that do not involve \(\dot{\bm{\beta}}\), minimising \(D_{KL}\) becomes
equal to finding \(\hat{\dot{\bm{\beta}}}\):

\[\hat{\dot{\bm{\beta}}} = \underset{\dot{\bm{\beta}}}{\operatorname{argmax}} \sum_{i=1}^n \left[ \bar{\pi}_i \{ \log(h(\bm{x}_i; \dot{\bm{\beta}})) + (1 - \bar{\pi}_i) \log(1 - h(\bm{x}_i; \dot{\bm{\beta}})) \} \right].\]

The summation has the structure of the log-likelihood for binary
responses, except that the responses are replaced by posterior means
from the main model. Most implementations of logistic regression can
maximise this likelihood by accepting posterior means as soft-label
responses, whose MLE-based estimate of \(\dot{\bm{\beta}}\) provides the
solution to this maximisation (e.g., \(\texttt{glm()}\) in R, which only
gives a warning that responses are not binary but continues to return
the MLEs). As such, by defining
\(h(\mathbf{x}_i, \dot{\bm{\beta}}) = (1 + e^{-\mathbf{x}_i^T \dot{\bm{\beta}}})^{-1}\),
we arrive at a second logistic equation, whose MLE-based predictions
approximate the posterior mean (replacing
\(\mathbb{E}[\text{logit}^{-1}(.)]\) with a single
\(\text{logit}^{-1}(.)\) evaluation at deployment).

This approximation can be used, for example, when one does not wish to
quantify deployment-time uncertainty, but still wishes to use the
posterior mean for decision-making (for the reasons explained above), or
when one wants to turn a model into a nomogram, or even simplify a
model. For example, a complex model that is implemented into an EMR
system can also be accompanied by a simpler nomogram based on a subset
of predictors. The link function can also be selected differently. For
example, an identity link results in an additive model on the
probability scale, which can further streamline nomogram calculations.

\subsection{Further considerations for summarising the posterior
distribution}\label{further-considerations-for-summarising-the-posterior-distribution}

A key appeal of the Bayesian framework is the intuitive presentation and
interpretation of prediction uncertainty. Posterior distributions are
generally accessible to clinicians and patients with minimal training,
and the entire distribution can be visualised to communicate
uncertainty\citep{zeger_bayescom_2020}. In a Bayesian framework, this
would involve a direct visualisation of the probability density of the
linear predictor transformed via the inverse link function (e.g., the
logit-normal distribution for logistic regression). For patient
communication, other intuitive measures of uncertainty can also be
considered. For example, Thomassen et al.~suggest reporting the
effective sample size as a measure of prediction
uncertainty\citep{thomassen_effectiven_2024}. The variance-matching
equations proposed by Thomassen et al.~are easily derivable at
deployment under this framework.

For this work, we focus on credible intervals as standard outputs of
Bayesian prediction, which can be intuitively interpreted as the
patient-specific range containing the outcome probability with a certain
coverage (often 95\%). The standard approach for summarising the
posterior distribution is to report the posterior mean as a measure of
centrality, coupled with tail-based intervals (e.g., the 2.5th to 97.5th
percentiles for 95\% coverage). Given the monotonic link function, the
percentiles of predictions correspond directly to the percentiles of
\(\eta\), with the latter being derived from the tail probabilities of
the normal distribution. This approach is practical and aligned with
typical Bayesian reporting (including the typical summarisation of draws
from posterior distributions in MCMC-based sampling); however, a few
alternatives deserve mention. One could construct intervals based on the
highest density regions of the posterior distribution, which generally
yield the shortest possible width. Further, while the posterior mean can
theoretically fall outside tail-based boundaries, this should be very
rare unless the SD of \(\eta\) is implausibly large. If so desired, one
could create an interval geometrically centered on the posterior mean.
However, both highest-density and mean-centered intervals require
deployment-time numerical methods and lack the standard status of
tail-based intervals. Therefore, we do not investigate them in this
study.

\section{Simulation studies to examine the Bayesian approaches and the
performance of posterior mean}\label{sec-simulations}

We conducted simulation studies to assess various aspects of the
proposed approach both compared with the plug-in predictions and
MCMC-based Bayesian approaches. The simulation design and reporting
generally follow the ADEMP framework \cite{morris_ademp_2019}.

\textbf{Aims}: These simulations are meant to test the key steps of the
proposed workflow: the regularising effect of the proposed priors,
normal approximation of the posterior distribution of predictor effects,
various deployment-time posterior mean computations, and the coverage of
the ensuing posterior distributions for individual-specific predicted
probabilities.

Two sets of simulations were conducted: (i) using data from a clinical
trial, and (ii) using fully synthetic data. The former is a `real-world'
example where we need not make any assumptions on the data-generation
mechanism, while the latter provides opportunities for changing major
features of the development sample such as outcome prevalence, as well
as studying the coverage of the posterior distribution (as we know the
true outcome probabilities).

\textbf{General methods}: All simulations were performed in R (version
4.6.0)\citep{R}. We computed predictions using flat, Jeffreys,
log-F(2,2), and Gaussian priors. We also examined two approaches as
general benchmarks: plug-in predictions from the frequentist ridge
regression and the posterior mean from a fully Bayesian ridge
regression. The former represents frequentist shrinkage approaches that
can be seen as the contemporary reference standards for model
development in small samples. The latter is a no-compromise Bayesian
approach without the approximation inherent in the Laplace method. We
used the \texttt{glmnet} package for the former and Stan (via
\texttt{rstan} package) for the latter\citep{stan}. For the frequentist
ridge regression, the tuning parameter was determined via
cross-validation, minimising deviance (with glmnet's default settings).
For the MCMC-based Bayesian ridge regression, it was assigned a half-t
distribution with 3 degrees of freedom, with 4 chains of 2,000
iterations each (1,000 warm-up), yielding 4,000 post-warm-up draws. We
also investigated both frequentist and MCMC-based Bayesian LASSO, but
they were not superior to their ridge counterparts and so are not
reported here to avoid clutter in the results.

\textbf{Targets}: Given the defence of the posterior mean under decision
theory, our primary emphasis was on clinical utility, which was assessed
in terms of standardised incremental NB (sNB, equal to the difference
between the NB of the model and the best default strategy, divided by
outcome prevalence in the sample). Of note, the sNB at treatment (risk)
threshold \(z\) can be defined as
\(sNB(z) = \mathbb{E}[I(\pi \ge z)(Y - (1-Y)z/(1-z))]/\mathbb{E}(Y) - \max(0, \mathbb{E}[Y-(1-Y)z/(1-z)]/\mathbb{E}(Y))\).

In addition to comparing sNBs across different priors, we studied the
difference in sNB with the use of posterior mean versus plug-in
predictions within the same setup (risk threshold, sample size, prior
family). To contextualise the magnitude of differences, we computed the
approximate sample size required for the plug-in prediction to achieve
the same expected sNB as the posterior mean. This was performed using
the inverse-power learning curve approach proposed by Figueroa et
al.\citep{figueroa_kearningcurve_2012}. The three parameters of this
curve were fitted to the four available degrees of freedom: the expected
NB values at the three sample sizes studied and the limit of sNB as
\(n \to \infty\). The latter was estimated from fitting the candidate
model on the entire population data.

We also assessed the performance of competing approaches in terms of
statistical measures, including prediction error (MSE), discrimination
(c-statistic), calibration. Results are summarised as mean \(\pm\) SD
for MSE and c-statistics, and median \(\pm\) IQR for calibration metrics
(O/E ratio and calibration slope), as the latter metrics can take large
values and their distribution might not have an expected value.

\subsection{Simulations using GUSTO
data}\label{simulations-using-gusto-data}

GUSTO-I was a clinical trial comparing different treatments for the
emergency management of heart attacks, and here we use the dataset to
develop prediction models for the outcome of 30-day
mortality\citep{noauthor_international_1993}. This study is widely used
for methodological research in clinical prediction
modelling\citep{ennis_comparison_1998, steyerberg_internal_2001, steyerberg_stepwise_1999, steyerberg_validation_2004, sadatsafavi_expected_2025}.
This binary outcome could be verified for all individuals.

\textbf{Data generating mechanism}: The full dataset is of size 40,830
(with 2,851 events), much larger than the sample sizes where the choice
of the prior and the posterior summary can make a difference. We used
subsets of this dataset, and used the held-out sample for performance
assessment without having to specify a true model. Our model for
predicting 30-day mortality will be similar to those used in previous
studies, consisting of seven
predictors\citep{sadatsafavi_expected_2025}. Table \ref{tab:gusto_desc}
shows the summary statistics of the seven candidate predictors and the
outcome.

\begin{table}[!h]
\centering
\caption{\label{tab:gusto_desc}Characteristics of the GUSTO-I development sample}
\centering
\fontsize{8}{10}\selectfont
\begin{tabular}[t]{lr}
\toprule
Variable & Summary (N = 40,830)\\
\midrule
Age (years) & 60.9 (11.9)\\
MI location: other (\%) & 1435 (3.5)\\
MI location: anterior (\%) & 15900 (38.9)\\
Previous MI (\%) & 6726 (16.5)\\
Killip class & I: 34825 (85.3\%); II: 5141 (12.6\%); III: 551 (1.3\%); IV: 313 (0.8\%)\\
Systolic BP, capped at 100 (mmHg) & 99.0 (4.6)\\
Pulse rate (beats/min) & 75.4 (17.8)\\
30-day mortality (\%) & 2851 (7.0)\\
\bottomrule
\multicolumn{2}{l}{\textsuperscript{} MI: Myocardial infarction; BP: Blood pressure}\\
\end{tabular}
\end{table}

\textbf{Methods:} We chose three sample sizes: 250, 500, and 1000. The
expected number of events are 17.5, 34.9, and 69.8. The sNBs in this
simulation were assessed at 2\%, 5\%, and 10\% thresholds. The number of
simulation runs was decided by the precision around difference in sNBs
between the posterior mean and plug-in predictions, which we consider as
the primary measure of interest. For each sample size, we run the
simulation 500 times (total simulation runs 1,500), after a pilot run
indicated that this will result in all pairwise comparison of sNBs
having a coefficient of variation (Monte Carlo SE of the difference
divided by its mean) \textless{} 1.96, thus being statistically
significant at the 0.05 level against Monte Carlo noise. This also
generally provided trivial Monte Carlo SE against variability for other
metrics.

\textbf{Results}: Table \ref{tab:gusto_nb} quantifies the clinical
utility of predictions: for each prior, training sample size, and
decision threshold, it reports the mean net benefit of the plug-in (PE)
and posterior-mean (PM) predictions and their difference (\(\Delta\)).
In general, the posterior mean outperformed the plug-in predictions
across all scenarios (with only in one scenario: flat prior at n=500 and
0.10 threshold, the difference not being significant against Monte Carlo
noise). At the 0.02 threshold, averaged across the four prior types, the
sNB of plug-in predictions was 0.0253, whereas for posterior mean it was
0.0303, which is a 19.5\% increase. The gain becomes more interpretable
when expressed as an equivalent development sample size, \(n^*\): the
size at which the plug-in attains the posterior mean's expected net
benefit. For example, under the Jeffreys prior, the plug-in predictions
should be developed using a sample size of approximately 327 to achieve
the expected NB of posterior mean predictions developed on \(n=250\).
This equivalent gain was largest under the flat prior and smallest under
the Bayesian ridge. As expected, the difference in efficiency between
the posterior mean and plug-in predictions diminished at larger sample
sizes.

\begin{table}[!h]
\centering
\caption{\label{tab:gusto_nb}GUSTO: standardised net benefit of posterior-mean (PM) versus plug-in (PE) predictions, by prior and training sample size, across decision thresholds}
\centering
\resizebox{\ifdim\width>\linewidth\linewidth\else\width\fi}{!}{
\fontsize{8}{10}\selectfont
\begin{tabular}[t]{rlrrrrrrrrrrrrrrr}
\toprule
\multicolumn{2}{c}{\textbf{ }} & \multicolumn{5}{c}{\textbf{$z = 0.02$}} & \multicolumn{5}{c}{\textbf{$z = 0.05$}} & \multicolumn{5}{c}{\textbf{$z = 0.10$}} \\
\cmidrule(l{3pt}r{3pt}){3-7} \cmidrule(l{3pt}r{3pt}){8-12} \cmidrule(l{3pt}r{3pt}){13-17}
$n$ & Prior & PE & PM & $\Delta$NB & CV & $n^*$ & PE & PM & $\Delta$NB & CV & $n^*$ & PE & PM & $\Delta$NB & CV & $n^*$\\
\midrule
250 & Flat & -0.0186 & 0.0217 & 0.0402 & 0.09 & 520 & 0.1919 & 0.2234 & 0.0314 & 0.09 & 369 & 0.2912 & 0.2981 & 0.0070 & 0.26 & 273\\
250 & Jeffreys & 0.0137 & 0.0212 & 0.0076 & 0.18 & 327 & 0.2157 & 0.2224 & 0.0067 & 0.21 & 279 & 0.3047 & 0.3085 & 0.0039 & 0.31 & 266\\
250 & log-F & 0.0071 & 0.0236 & 0.0165 & 0.14 & 401 & 0.2114 & 0.2327 & 0.0213 & 0.11 & 356 & 0.3040 & 0.3215 & 0.0175 & 0.10 & 338\\
250 & Bayesian ridge & 0.0087 & 0.0139 & 0.0051 & 0.28 & 295 & 0.2015 & 0.2099 & 0.0084 & 0.24 & 278 & 0.2896 & 0.3054 & 0.0159 & 0.10 & 308\\
250 & Ridge (CV-min) & 0.0081 &  &  &  &  & 0.1937 &  &  &  &  & 0.2832 &  &  &  & \\
250 & Bayesian ridge (MCMC) &  & 0.0143 &  &  &  &  & 0.2125 &  &  &  &  & 0.2991 &  &  & \\
500 & Flat & 0.0190 & 0.0315 & 0.0125 & 0.08 & 836 & 0.2356 & 0.2458 & 0.0102 & 0.07 & 639 & 0.3319 & 0.3325 & 0.0006 & 1.09 & 507\\
500 & Jeffreys & 0.0293 & 0.0310 & 0.0017 & 0.25 & 563 & 0.2443 & 0.2480 & 0.0037 & 0.14 & 562 & 0.3376 & 0.3420 & 0.0043 & 0.13 & 568\\
500 & log-F & 0.0287 & 0.0324 & 0.0036 & 0.13 & 627 & 0.2436 & 0.2508 & 0.0071 & 0.08 & 626 & 0.3372 & 0.3446 & 0.0073 & 0.08 & 624\\
500 & Bayesian ridge & 0.0243 & 0.0263 & 0.0019 & 0.28 & 559 & 0.2394 & 0.2449 & 0.0056 & 0.12 & 580 & 0.3315 & 0.3393 & 0.0078 & 0.08 & 605\\
500 & Ridge (CV-min) & 0.0220 &  &  &  &  & 0.2380 &  &  &  &  & 0.3293 &  &  &  & \\
500 & Bayesian ridge (MCMC) &  & 0.0269 &  &  &  &  & 0.2430 &  &  &  &  & 0.3349 &  &  & \\
1000 & Flat & 0.0347 & 0.0378 & 0.0031 & 0.17 & 1392 & 0.2596 & 0.2628 & 0.0032 & 0.10 & 1216 & 0.3546 & 0.3560 & 0.0014 & 0.28 & 1067\\
1000 & Jeffreys & 0.0367 & 0.0375 & 0.0008 & 0.25 & 1132 & 0.2618 & 0.2633 & 0.0014 & 0.16 & 1113 & 0.3567 & 0.3590 & 0.0023 & 0.10 & 1134\\
1000 & log-F & 0.0370 & 0.0378 & 0.0008 & 0.20 & 1120 & 0.2620 & 0.2640 & 0.0020 & 0.11 & 1161 & 0.3567 & 0.3594 & 0.0027 & 0.09 & 1157\\
1000 & Bayesian ridge & 0.0343 & 0.0349 & 0.0006 & 0.45 & 1078 & 0.2600 & 0.2617 & 0.0017 & 0.16 & 1114 & 0.3547 & 0.3576 & 0.0029 & 0.08 & 1148\\
1000 & Ridge (CV-min) & 0.0326 &  &  &  &  & 0.2595 &  &  &  &  & 0.3541 &  &  &  & \\
1000 & Bayesian ridge (MCMC) &  & 0.0353 &  &  &  &  & 0.2611 &  &  &  &  & 0.3559 &  &  & \\
\bottomrule
\multicolumn{17}{l}{\textsuperscript{} PE: plug-in (point-estimate) prediction; PM: posterior-mean prediction; $\Delta$NB = NB(PM) $-$ NB(PE); CV = Monte Carlo SE\,/\,$|{\Delta}\text{NB}|$. $n^*$: plug-in}\\
\multicolumn{17}{l}{development size whose net benefit equals the posterior mean's at this $n$, from an inverse-power learning curve $\text{NB}(n) = \text{NB}_{\infty} - b\,n^{c}$, where}\\
\multicolumn{17}{l}{$\text{NB}_{\infty}$ is the net benefit of the candidate model fitted to the full cohort and $b$ and $c$ are fitted to the three simulated sizes. The frequentist ridge}\\
\multicolumn{17}{l}{(CV-min) and the MCMC Bayesian ridge are single-prediction comparators (plug-in and posterior mean, respectively); their net benefit is shown in the PE or PM column and the}\\
\multicolumn{17}{l}{remaining cells are left blank as there is no PE-vs-PM contrast.}\\
\end{tabular}}
\end{table}

\begin{figure}[!htbp]
  \centering
  \caption{Comparison between plug-in (PE) and posterior mean (PM) estimates of predicted probabilities across various priors in GUSTO simulations across different sample sizes}
  \includegraphics[width=0.9\linewidth,height=0.75\textheight,keepaspectratio]{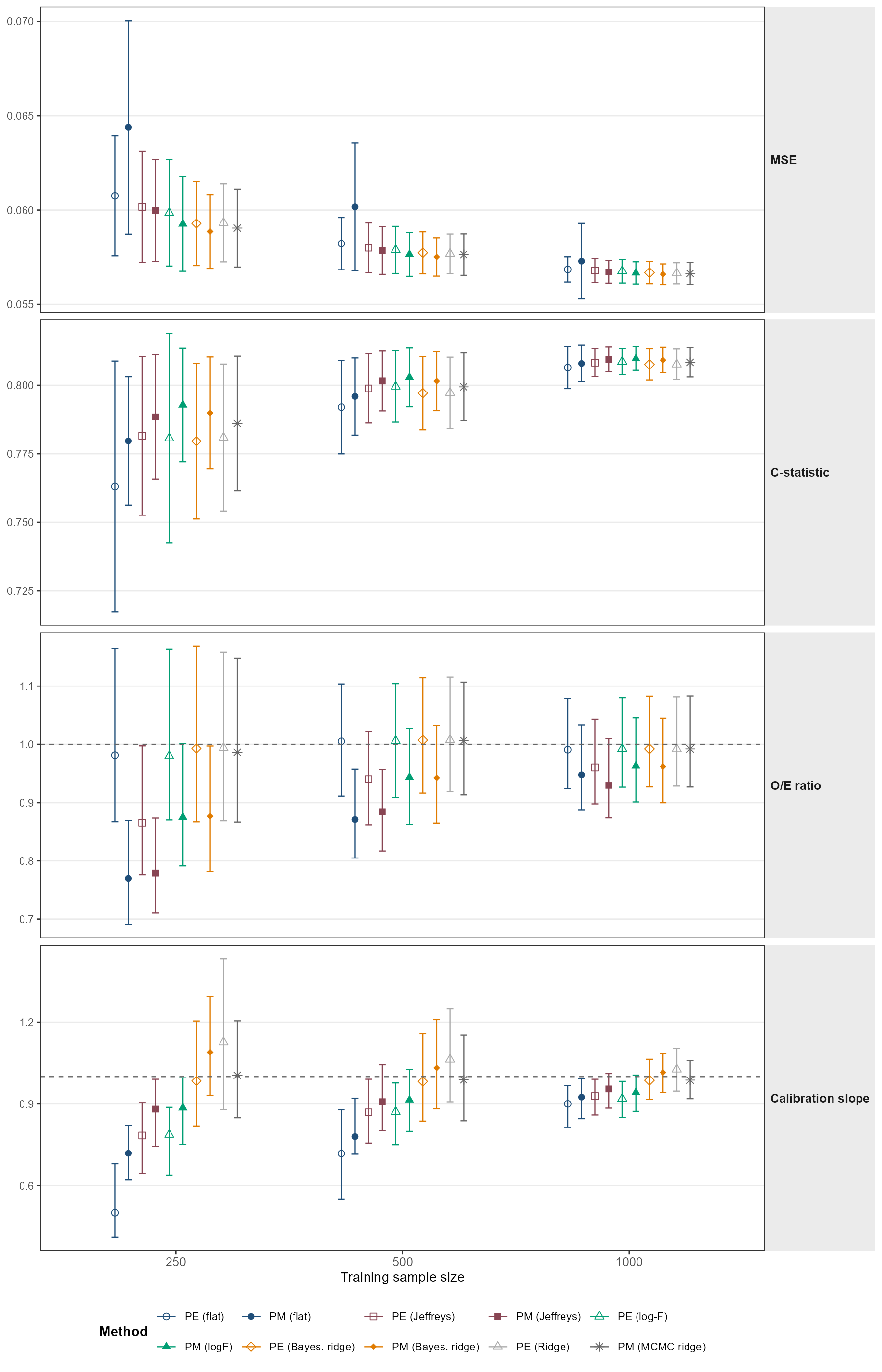}
  \caption*{\scriptsize EPV: Events per variable; MSE: mean squared error; O/E: Observed-to-expected ratio; sNB: Standardised net benefit}
  \caption*{\scriptsize Points and whiskers represent mean and SD for MSE and c-statistic, and median and 25-75\% quantile range for O/E ratio and calibration slope}
  \label{fig:sim_gusto_main}
\end{figure}

Results for statistical performance metrics are provided in Figure
\ref{fig:sim_gusto_main}. An immediate observation is the high
variability of the flat-prior model in small samples, manifested in wide
SDs for the MSE of the posterior mean, and for c-statistic for the
plug-in predictions. For the flat prior at small samples, posterior mean
had higher MSE than the plug-in predictions. Flat-prior and Bayesian
ridge plug-in predictions, as well as frequentist and MCMC ridge,
resulted in O/E\textasciitilde1 (as expected). The plug-in predictions
based on flat prior had a calibration slope far away from the ideal
value of 1. The posterior mean nudges O/E ratio away from 1, but
improves the slope.

\begin{figure}[!htbp]
  \centering
  \caption{Comparison between different posterior mean (PM) computation methods in GUSTO simulations}
  \includegraphics[width=0.9\linewidth,height=0.75\textheight,keepaspectratio]{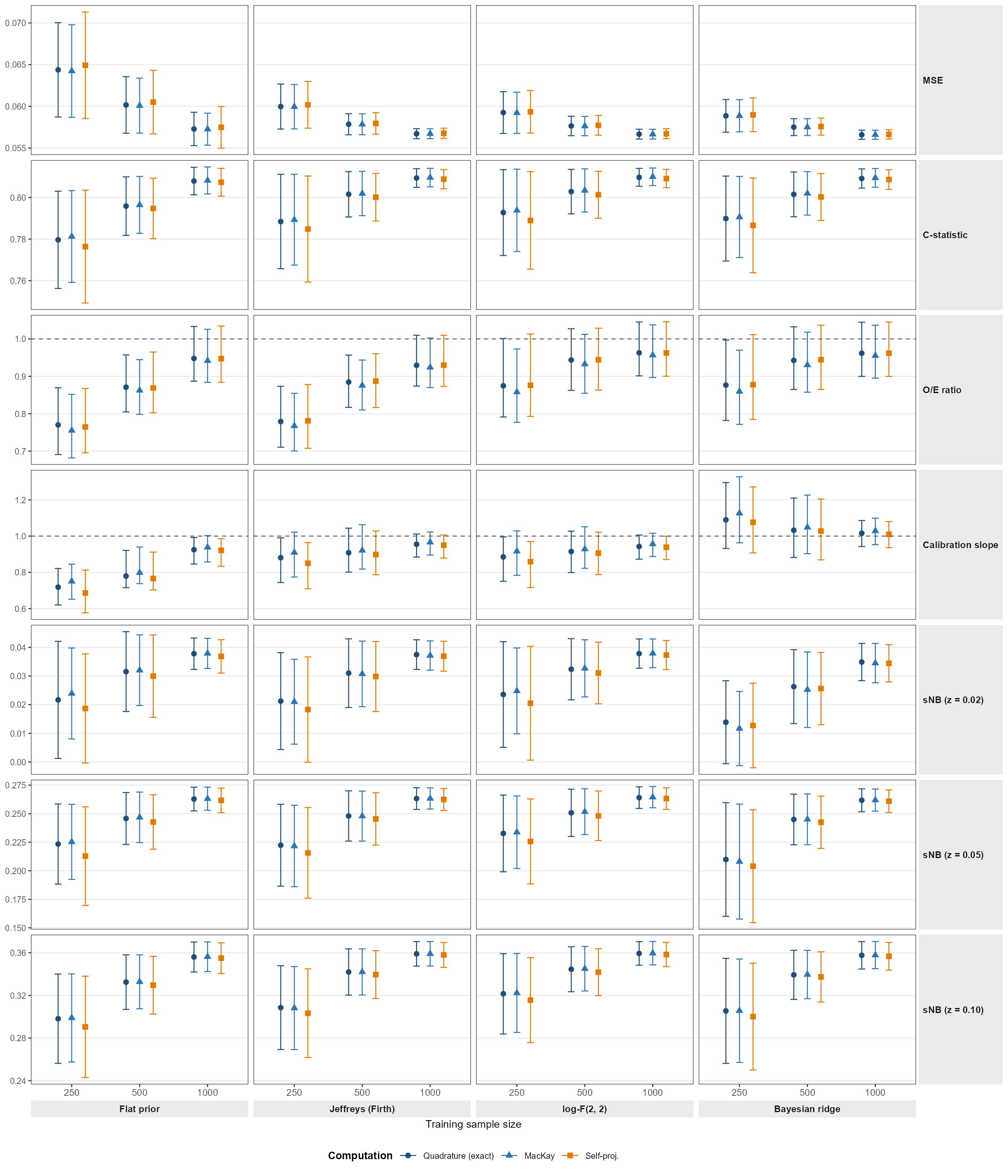}
  \caption*{\scriptsize EPV: Events per variable; MSE: mean square error; O/E: Observed-to-expected ratio; sNB: Standardised net benefit}
  \caption*{\scriptsize Points and whiskers represent mean and SD for MSE and c-statistic, and median and 25-75\% quantile range for O/E ratio and calibration slope}
  \label{fig:sim_gusto_pm}
\end{figure}

The comparison between various computation methods is provided in Figure
\ref{fig:sim_gusto_pm}. Compared to the exact (quadrature) method,
prediction by MacKay approximation had nearly identical performance.
Interestingly, while in several scenarios it slightly reduced the
performance of the posterior mean compared with the exact method, in
some scenarios it improved it. This might be due to the approximation
error acting in the opposite direction of model misspecification error.
The two-stage projections, on the other hand, generally resulted in some
loss of performance.

\subsection{Simulations using synthetic
data}\label{simulations-using-synthetic-data}

\textbf{Data generating mechanism}: The sample sizes varied to generate
expected events per variable (EPV) of 2.5, 5, and 10. We considered
three types of models: with 5, 10, and 30 predictors, representing
simple, average, and complex models. Within each simulation, this model
structure was applied separately to both the true model and the
candidate model. Outcome probability was also varied: 5\%, 15\%, and
30\%. We chose a fully factorial design across the afore-mentioned 4
categories (EPV, size of candidate model, size of the true model, and
outcome prevalence) - creating 81 unique designs. Within each design,
the following aspects were randomly varied: predictors could have a 50\%
chance of being continuous or binary, and predictors could randomly have
correlation coefficients of 0, 0.25, 0.50, and 0.75. Within each
simulation, true coefficients were independently sampled from a double
exponential (Laplace) distribution with zero mean and SD of 0.5. We
chose this distribution because the normal distribution typically used
in simulation studies will provide an unfair advantage to the methods
based on Gaussian priors. The intercept was then determined numerically
to match the target prevalence.

\textbf{Methods:} Because the outcome prevalence is widely varying in
this setup, we computed sNBs at three relative thresholds: at 25th, 50th
and 75th percentiles of the distribution of true probabilities in the
population. We used two strategies to reduce Monte Carlo variance in
population-level metric estimates. First, population covariate vectors
were generated using a quasi-random approach (Sobol sequence) rather
than independent random draws. This scheme can achieve convergence close
to \(O(n^{-1})\), compared with \(O(n^{-1/2})\) for independent
sampling\citep{caflisch_monte_1998}. Second, all metrics were assessed
against the simulated true probabilities \(p\) rather than the simulated
response \(Y\), thereby integrating out the Bernoulli variation (e.g.,
for MSE, \(\mathbb{E}[(\pi_i -Y_i)^2]=p_i(1-\pi_i)^2+(1-p_i)\pi_i^2\),
and using the latter expression lowers Monte Carlo variance). Pilot runs
indicated that a population size of \(10^4\) was sufficient under these
strategies to render population-level Monte Carlo noise negligible.
Given the smaller difference in sNB (and other metric) performances, the
pilot assessment indicated that 1,500 simulation runs for each EPV would
be required to achieve a similar CV to the GUSTO-based simulations
(resulting in 4,500 unique simulations in total).

\textbf{Results}: Differences in expected sNBs and sample size
equivalents (\(EPV^*\)) are presented in Table \ref{tab:synth_snb}.
Compared to GUSTO-based simulations, differences in sNBs were less
pronounced in synthetic simulations, and mostly discernible at the lower
thresholds. Again, the efficiency of prediction approaches can be
expressed as an equivalent events-per-variable, EPV\(^*\) : the plug-in
EPV at which the plug-in predictions attain the posterior mean's
expected sNB. Under the Jeffreys prior, for example, at EPV=2.5 and the
low threshold, the plug-in predictor would need EPV\(^* \approx\) 3.9 to
match the posterior mean developed at EPV=2.5 in expected clinical
utility. This represents a 56.2\% efficiency gain for the posterior
mean. The analogous gain was 14.5\% at the middle threshold. On the
other hand, there was a small loss of clinical utility at the highest
threshold when using posterior mean instead of plug-in predictions.

\begin{table}[!h]
\centering
\caption{\label{tab:synth_snb}Synthetic simulations: standardised net benefit of posterior-mean (PM) versus plug-in (PE) predictions, by prior and EPV, across decision thresholds}
\centering
\resizebox{\ifdim\width>\linewidth\linewidth\else\width\fi}{!}{
\fontsize{8}{10}\selectfont
\begin{tabular}[t]{rlrrrrrrrrrrrrrrr}
\toprule
\multicolumn{2}{c}{\textbf{ }} & \multicolumn{5}{c}{\textbf{$z = 0.25$}} & \multicolumn{5}{c}{\textbf{$z = 0.50$}} & \multicolumn{5}{c}{\textbf{$z = 0.75$}} \\
\cmidrule(l{3pt}r{3pt}){3-7} \cmidrule(l{3pt}r{3pt}){8-12} \cmidrule(l{3pt}r{3pt}){13-17}
EPV & Prior & PE & PM & $\Delta$sNB & CV & EPV$^*$ & PE & PM & $\Delta$sNB & CV & EPV$^*$ & PE & PM & $\Delta$sNB & CV & EPV$^*$\\
\midrule
2.5 & Flat & -0.0362 & -0.0139 & 0.0223 & 0.05 & 4.1 & 0.0455 & 0.0582 & 0.0127 & 0.07 & 3.2 & 0.0950 & 0.0823 & -0.0127 & 0.13 & 2.1\\
2.5 & Jeffreys & -0.0165 & -0.0053 & 0.0111 & 0.05 & 3.9 & 0.0574 & 0.0624 & 0.0051 & 0.11 & 2.9 & 0.0900 & 0.0810 & -0.0090 & 0.07 & 2.2\\
2.5 & log-F & -0.0245 & -0.0103 & 0.0142 & 0.05 & 3.8 & 0.0523 & 0.0613 & 0.0091 & 0.07 & 3.1 & 0.0998 & 0.0945 & -0.0053 & 0.11 & 2.3\\
2.5 & Bayesian ridge & -0.0066 & -0.0015 & 0.0051 & 0.14 & 3.3 & 0.0523 & 0.0553 & 0.0030 & 0.25 & 2.7 & 0.0997 & 0.0959 & -0.0038 & 0.18 & 2.3\\
2.5 & Ridge (CV-min) & -0.0050 &  &  &  &  & 0.0527 &  &  &  &  & 0.1003 &  &  &  & \\
2.5 & Bayesian ridge (MCMC) &  & -0.0057 &  &  &  &  & 0.0568 &  &  &  &  & 0.1046 &  &  & \\
5.0 & Flat & -0.0069 & -0.0005 & 0.0064 & 0.06 & 6.5 & 0.0731 & 0.0769 & 0.0038 & 0.10 & 5.6 & 0.1242 & 0.1212 & -0.0029 & 0.15 & 4.6\\
5.0 & Jeffreys & -0.0008 & 0.0033 & 0.0041 & 0.08 & 6.5 & 0.0767 & 0.0786 & 0.0019 & 0.16 & 5.4 & 0.1217 & 0.1173 & -0.0044 & 0.09 & 4.5\\
5.0 & log-F & -0.0040 & 0.0010 & 0.0049 & 0.06 & 6.4 & 0.0744 & 0.0777 & 0.0033 & 0.09 & 5.6 & 0.1255 & 0.1229 & -0.0026 & 0.12 & 4.7\\
5.0 & Bayesian ridge & 0.0040 & 0.0055 & 0.0015 & 0.19 & 5.7 & 0.0721 & 0.0727 & 0.0007 & 0.58 & 5.1 & 0.1259 & 0.1246 & -0.0013 & 0.43 & 4.8\\
5.0 & Ridge (CV-min) & 0.0044 &  &  &  &  & 0.0717 &  &  &  &  & 0.1255 &  &  &  & \\
5.0 & Bayesian ridge (MCMC) &  & 0.0042 &  &  &  &  & 0.0753 &  &  &  &  & 0.1276 &  &  & \\
10.0 & Flat & 0.0066 & 0.0085 & 0.0020 & 0.08 & 11.6 & 0.0918 & 0.0931 & 0.0014 & 0.13 & 10.8 & 0.1471 & 0.1459 & -0.0012 & 0.15 & 9.5\\
10.0 & Jeffreys & 0.0085 & 0.0100 & 0.0015 & 0.08 & 11.7 & 0.0930 & 0.0941 & 0.0011 & 0.19 & 10.7 & 0.1458 & 0.1442 & -0.0016 & 0.15 & 9.3\\
10.0 & log-F & 0.0074 & 0.0092 & 0.0018 & 0.07 & 11.7 & 0.0921 & 0.0934 & 0.0013 & 0.13 & 10.8 & 0.1476 & 0.1465 & -0.0012 & 0.15 & 9.4\\
10.0 & Bayesian ridge & 0.0109 & 0.0113 & 0.0004 & 0.33 & 10.6 & 0.0905 & 0.0908 & 0.0003 & 0.72 & 10.2 & 0.1486 & 0.1481 & -0.0005 & 0.48 & 9.8\\
10.0 & Ridge (CV-min) & 0.0108 &  &  &  &  & 0.0897 &  &  &  &  & 0.1476 &  &  &  & \\
10.0 & Bayesian ridge (MCMC) &  & 0.0108 &  &  &  &  & 0.0919 &  &  &  &  & 0.1488 &  &  & \\
\bottomrule
\multicolumn{17}{l}{\textsuperscript{} EPV: events per variable; threshold $z$ is the given quantile of the true-risk distribution; PE: plug-in (point-estimate) prediction; PM: posterior-mean prediction; $\Delta$sNB =}\\
\multicolumn{17}{l}{sNB(PM) $-$ sNB(PE); CV = Monte Carlo SE\,/\,$|{\Delta}\text{sNB}|$. EPV$^*$: plug-in events-per-variable whose sNB equals the posterior mean's at this EPV, from an}\\
\multicolumn{17}{l}{inverse-power learning curve $\text{sNB}(\text{EPV}) = \text{sNB}_{\infty} - b\,\text{EPV}^{c}$, where $\text{sNB}_{\infty}$ is the sNB of the candidate model at its}\\
\multicolumn{17}{l}{pseudo-true coefficients and $b$ and $c$ are fitted to the three EPV levels; EPV$^* <$ EPV indicates PM underperforms PE there. The frequentist ridge (CV-min) and the MCMC Bayesian}\\
\multicolumn{17}{l}{ridge are single-prediction comparators; their sNB is shown in the PE or PM column respectively and the remaining cells are blank.}\\
\end{tabular}}
\end{table}

\begin{figure}[!htbp]
  \centering
  \caption{Comparison between plug-in (PE) and posterior mean (PM) estimates of predicted probabilities across various priors in synthetic simulations}
  \includegraphics[width=0.95\linewidth]{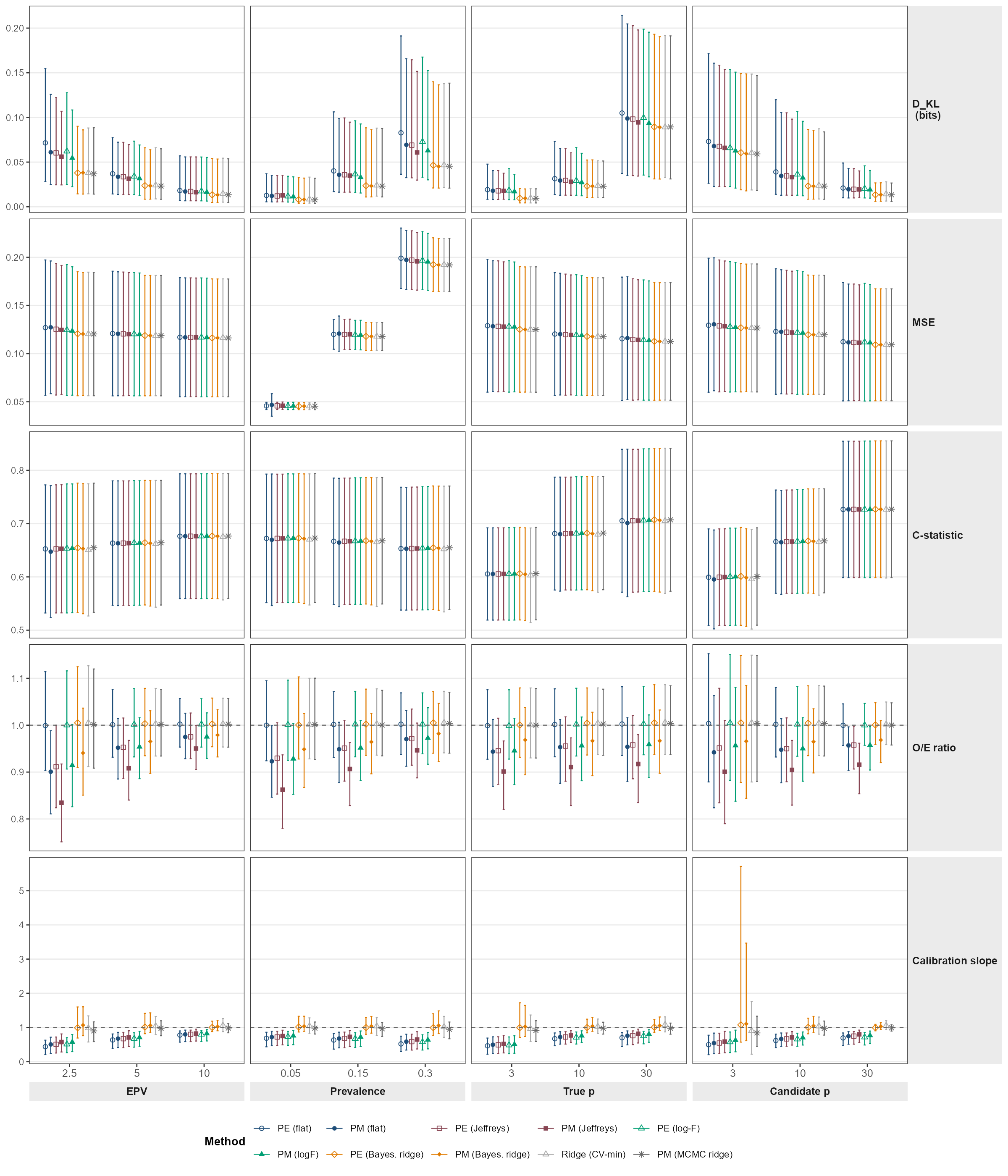}
  \caption*{\scriptsize EPV: Events per variable; MSE: mean squared error; O/E: Observed-to-expected ratio; sNB: Standardised net benefit}
  \caption*{\scriptsize Points and whiskers represent mean and SD for MSE and c-statistic, and median and 25-75\% quantile range for O/E ratio and calibration slope}
  \label{fig:sim_synth_main}
\end{figure}

Results for statistical measures of performance are provided in Figure
\ref{fig:sim_synth_main}. The information loss (\(D_{KL}\)) was lower
for the posterior mean compared with plug-in predictions in small
samples. Differences in MSE and c-statistics were small across methods.
The trade-off in O/E and calibration slope between the plug-in
prediction and posterior mean is visible especially with small EPVs.
Again, with small EPVs, the posterior mean reduced the O/E ratio away
from 1 (by \textasciitilde0.05), but improved calibration slope towards
1 (by a similar amount). Posterior mean with Jeffreys prior generated
the smallest O/Es (albeit the median O/E remained \textgreater0.9) -
this reflects the compounded effect of the shrinkage-to-0.5 of both the
prior distribution and the posterior mean. With candidate models of
small sizes, the Bayesian and frequentist ridge shrank predictor effects
to close to 0 in many scenarios, resulting in a model with nearly flat
predictions, reflecting in large calibration slope values. This is a
real possibility outside simulations when hierarchical methods are
nested within bootstrap resampling, especially in small samples where
shrinkage is needed the most. Aside from this anomaly, in several
scenarios the hierarchical shrinkage approaches (both Bayesian and
frequentist) in these scenarios performed somewhat better than simpler
priors. This can be considered an expected finding given that the true
regression coefficients have equal variances across populations, which
is compatible with the universal shrinkage enforced by these methods.

\begin{figure}[!htbp]
  \centering
  \caption{Comparison between different posterior mean (PM) computation methods in GUSTO simulations}
  \includegraphics[width=0.9\linewidth,height=0.75\textheight,keepaspectratio]{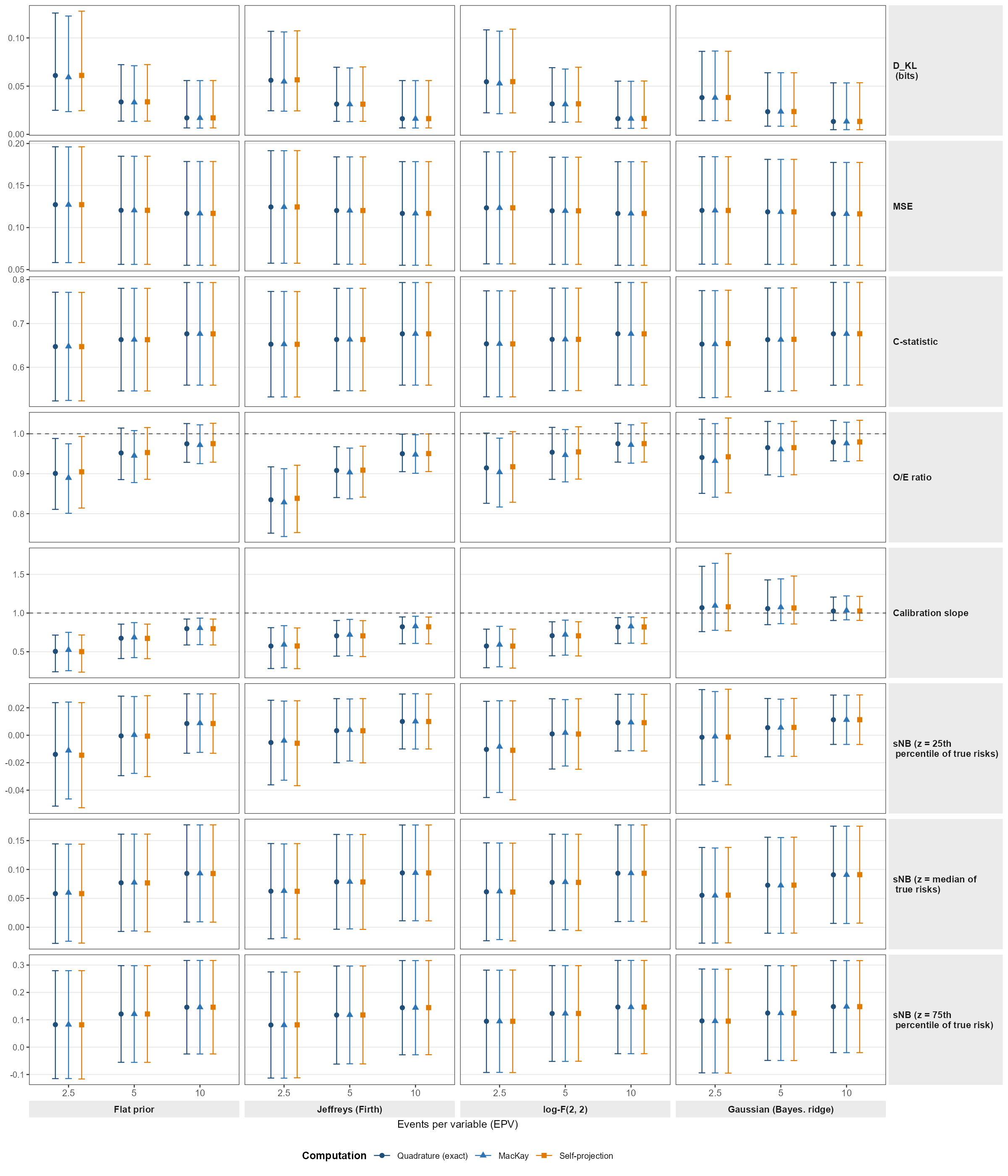}
  \caption*{\scriptsize EPV: Events per variable; MSE: mean square error; O/E: Observed-to-expected ratio; sNB: Standardised net benefit}
  \caption*{\scriptsize Points and whiskers represent mean and SD for MSE and c-statistic, and median and 25-75\% quantile range for O/E ratio and calibration slope}
  \label{fig:sim_synth_pm}
\end{figure}

The comparison among different posterior mean computation methods is
presented in Figure \ref{fig:sim_synth_pm}. Here, the difference among
methods is not as pronounced as in the previous simulation.

\subsubsection{Evaluation of the calibration of the individual posterior
distributions of outcome
probability}\label{evaluation-of-the-calibration-of-the-individual-posterior-distributions-of-outcome-probability}

We assessed whether the individual posterior distributions of outcome
probability properly capture uncertainty in model predictions; we refer
to this concept as checking the calibration of individual posterior
distributions. Whereas frequentist approaches to calibration assessment
focus on a model's point predictions, here in the Bayesian setting the
calibration focus is rather on the entire posterior distribution. As our
simulations cover both misspecified and correctly specified candidate
models, the true value (estimand) for each individual's predicted
probability should be carefully chosen. Under correct specification,
\(\bm{\beta}\)s converge to their true values, and the posterior
distribution of outcome probability concentrates around the true
conditional probability for each individual. Under model
misspecification, however, regression coefficients in the candidate
model space do not have a population counterpart and therefore do not
converge to any true value as the sample size increases. Instead, under
standard regularity conditions, they converge to `pseudo-true' values:
the unique set of parameter values within the candidate model space that
minimises the Kullback--Leibler divergence from the true
response-generating mechanism\citep{white_pseudo_1982}. Consequently,
the posterior distribution concentrates around the predictions implied
by these pseudo-true coefficients. Following the same logic as presented
in Section \ref{sec-projpred}, these pseudo-true coefficients can be
obtained from a soft-label logistic model. We fitted such a model
regressing true probabilities on the candidate predictors using the
entire simulated population.

Posterior distributions were assessed in two ways. First, we quantified
the actual coverage probability of the central (tail-based) 95\%
credible interval, computing the proportion of 95\% CrIs across
individuals that contained their (pseudo-)true probabilities. Secondly,
we note that if the posterior distribution is `calibrated', the
(pseudo-)true probability is a random draw from the posterior
distribution for each individual. This means
\(\Phi(\frac{\mbox{logit}(p_i)-\mu_i}{\sigma_i}) \sim \mbox{Uniform}(0,1)\),
where \(p_i\), \(\mu_i\), and \(\sigma_i\) are the (pseudo-)true
probabilities and mean and SD of the linear term \(\eta\) for the ith
individual, respectively, and \(\Phi\) is the standard normal CDF.
Hence, we assessed if the empirical distribution of this quantity
matches the uniform CDF, evaluated at three reference quantiles of 0.1,
0.5, and 0.9.

Results, stratified by EPV, are provided in Table
\ref{tab:posterior_calibration}. All methods generated acceptable
coverage probabilities, with the Bayesian ridge values slightly lower
than the nominal ones. The empirical CDFs also generally seem to follow
the uniform distribution, but they were slightly lower than expected for
flat and log-F distributions (indicating a slight shift of the posterior
distribution to the left), and higher than expected for the Bayesian
ridge at 0.1 and 0.5 (but not 0.9) thresholds - indicating a
narrower-than-expected spread of the posterior distribution.

\begin{table}[!h]
\centering
\caption{\label{tab:posterior_calibration}Posterior distribution assessment by prior and events-per-variable (EPV). Each cell shows mean (SD) across replicates. }
\centering
\fontsize{8}{10}\selectfont
\begin{tabular}[t]{lrrrrr}
\toprule
Prior & EPV & Coverage of 95\% CrI & $\hat{CDF}(0.10) ^a$ & $\hat{CDF}(0.50) ^a$ & $\hat{CDF}(0.90) ^a$\\
\midrule
\cellcolor{gray!10}{Flat} & \cellcolor{gray!10}{2.5} & \cellcolor{gray!10}{0.945 (0.068)} & \cellcolor{gray!10}{0.082 (0.085)} & \cellcolor{gray!10}{0.415 (0.170)} & \cellcolor{gray!10}{0.861 (0.122)}\\
Flat & 5.0 & 0.944 (0.074) & 0.089 (0.095) & 0.444 (0.169) & 0.875 (0.114)\\
\cellcolor{gray!10}{Flat} & \cellcolor{gray!10}{10.0} & \cellcolor{gray!10}{0.948 (0.066)} & \cellcolor{gray!10}{0.083 (0.084)} & \cellcolor{gray!10}{0.454 (0.165)} & \cellcolor{gray!10}{0.880 (0.110)}\\
\addlinespace
Jeffreys & 2.5 & 0.955 (0.059) & 0.113 (0.098) & 0.518 (0.171) & 0.916 (0.089)\\
\cellcolor{gray!10}{Jeffreys} & \cellcolor{gray!10}{5.0} & \cellcolor{gray!10}{0.947 (0.075)} & \cellcolor{gray!10}{0.115 (0.107)} & \cellcolor{gray!10}{0.517 (0.167)} & \cellcolor{gray!10}{0.913 (0.094)}\\
Jeffreys & 10.0 & 0.952 (0.062) & 0.104 (0.095) & 0.508 (0.165) & 0.907 (0.094)\\
\addlinespace
\cellcolor{gray!10}{log-F} & \cellcolor{gray!10}{2.5} & \cellcolor{gray!10}{0.954 (0.058)} & \cellcolor{gray!10}{0.078 (0.088)} & \cellcolor{gray!10}{0.434 (0.184)} & \cellcolor{gray!10}{0.877 (0.113)}\\
log-F & 5.0 & 0.949 (0.071) & 0.089 (0.098) & 0.454 (0.174) & 0.886 (0.108)\\
\cellcolor{gray!10}{log-F} & \cellcolor{gray!10}{10.0} & \cellcolor{gray!10}{0.952 (0.062)} & \cellcolor{gray!10}{0.084 (0.087)} & \cellcolor{gray!10}{0.461 (0.169)} & \cellcolor{gray!10}{0.888 (0.104)}\\
\addlinespace
Gaussian (Bayes. ridge) & 2.5 & 0.919 (0.130) & 0.139 (0.150) & 0.529 (0.218) & 0.910 (0.122)\\
\cellcolor{gray!10}{Gaussian (Bayes. ridge)} & \cellcolor{gray!10}{5.0} & \cellcolor{gray!10}{0.917 (0.126)} & \cellcolor{gray!10}{0.138 (0.141)} & \cellcolor{gray!10}{0.533 (0.189)} & \cellcolor{gray!10}{0.905 (0.114)}\\
Gaussian (Bayes. ridge) & 10.0 & 0.932 (0.104) & 0.124 (0.124) & 0.521 (0.183) & 0.907 (0.103)\\
\addlinespace
\cellcolor{gray!10}{MCMC ridge (exact)} & \cellcolor{gray!10}{2.5} & \cellcolor{gray!10}{0.950 (0.077)} & \cellcolor{gray!10}{0.084 (0.108)} & \cellcolor{gray!10}{0.461 (0.224)} & \cellcolor{gray!10}{0.877 (0.136)}\\
MCMC ridge (exact) & 5.0 & 0.941 (0.084) & 0.097 (0.115) & 0.480 (0.196) & 0.885 (0.120)\\
\cellcolor{gray!10}{MCMC ridge (exact)} & \cellcolor{gray!10}{10.0} & \cellcolor{gray!10}{0.951 (0.068)} & \cellcolor{gray!10}{0.091 (0.099)} & \cellcolor{gray!10}{0.480 (0.184)} & \cellcolor{gray!10}{0.890 (0.110)}\\
\bottomrule
\multicolumn{6}{l}{\textsuperscript{} CrI: Credible interval; EPV: Events per variable; CDF: Cumulative distribution function}\\
\multicolumn{6}{l}{\textsuperscript{} $^a$ For $\hat{CDF}(x)$, the desired value is $x$}\\
\end{tabular}
\end{table}

\section{An illustrative example}\label{an-illustrative-example}

To showcase the proposed approaches, we use an example based on GUSTO-I
data. We use a random subset of 500 individuals (with 27 events). We
adopt the same model equation as used in GUSTO-based simulations. We
fitted models with each of the proposed priors, as well as two
comparators: CV-tuned frequentist ridge, and fully Bayesian MCMC ridge.
With seven predictor parameters, there are 3.86 events per variable
(EPV) in the development sample, which is below the sample size
generally considered enough to minimise
overfitting\citep{riley_epv_2018}. Thus, we expect the choice of prior
and the impact of using the posterior mean rather than plug-in
prediction to be noticeable.

Parameter estimates for the various modelling approaches are provided in
Table \ref{tab:gusto_coefs}. The bottom rows of this Table show the
predictions for our exemplary patient Sam: a 61-year old individual with
an acute inferior MI, no previous MI history, with Killip class of I (a
measure of the severity of heart failure), pulse rate of 100, and
systolic blood pressure of 83. Predictions for Sam, including 95\% CrIs
when applicable, are provided at the bottom of this Table. For each
prior, in addition to the exact PM computation, we also compute the PM
using the MacKay's approximation and via self-projections.

\begin{table}[!h]
\centering
\begin{threeparttable}
\caption{\label{tab:gusto_coefs}Coefficient estimates (with standard errors where available) for each prior / method, and the corresponding plug-in estimate (PE), posterior mean (PM), and 95\% credible interval (CrI) for the exemplary patient.}
\centering
\fontsize{8}{10}\selectfont
\begin{tabular}[t]{>{\raggedright\arraybackslash}p{2.6cm}>{\raggedleft\arraybackslash}p{0.85cm}>{\raggedleft\arraybackslash}p{0.85cm}>{\raggedleft\arraybackslash}p{0.85cm}>{\raggedleft\arraybackslash}p{0.85cm}>{\raggedleft\arraybackslash}p{0.85cm}>{\raggedleft\arraybackslash}p{0.85cm}>{\raggedleft\arraybackslash}p{0.85cm}>{\raggedleft\arraybackslash}p{0.85cm}>{\raggedleft\arraybackslash}p{0.85cm}>{\raggedleft\arraybackslash}p{0.85cm}>{\raggedleft\arraybackslash}p{0.85cm}}
\toprule
\multicolumn{1}{c}{\textbf{ }} & \multicolumn{2}{c}{\textbf{\makecell[c]{Flat\\(MLE)}}} & \multicolumn{2}{c}{\textbf{Jeffreys}} & \multicolumn{2}{c}{\textbf{log-F(2,2)}} & \multicolumn{2}{c}{\textbf{\makecell[c]{Gaussian\\(Bayes. ridge)}}} & \multicolumn{1}{c}{\textbf{\makecell[c]{Ridge\\(CV-min)*}}} & \multicolumn{2}{c}{\textbf{\makecell[c]{Bayes. ridge\\(MCMC)}}} \\
\cmidrule(l{3pt}r{3pt}){2-3} \cmidrule(l{3pt}r{3pt}){4-5} \cmidrule(l{3pt}r{3pt}){6-7} \cmidrule(l{3pt}r{3pt}){8-9} \cmidrule(l{3pt}r{3pt}){10-10} \cmidrule(l{3pt}r{3pt}){11-12}
Term & Est. & SE & Est. & SE & Est. & SE & Est. & SE & Est. & Est. & SE\\
\midrule
(Intercept) & -4.187 & 4.010 & -3.654 & 3.851 & -4.244 & 3.954 & -2.878 & 3.537 & -2.236 & -3.213 & 3.673\\
Age & 0.069 & 0.022 & 0.066 & 0.021 & 0.069 & 0.022 & 0.047 & 0.019 & 0.038 & 0.052 & 0.020\\
MI location - other & 1.169 & 0.908 & 1.236 & 0.863 & 0.812 & 0.803 & 0.888 & 0.816 & 0.757 & 0.865 & 0.855\\
MI location - anterior & 0.693 & 0.496 & 0.660 & 0.471 & 0.563 & 0.454 & 0.511 & 0.402 & 0.420 & 0.550 & 0.420\\
Previous MI & 0.965 & 0.467 & 0.944 & 0.449 & 0.872 & 0.448 & 0.749 & 0.427 & 0.635 & 0.778 & 0.429\\
Killip class & 0.707 & 0.302 & 0.655 & 0.293 & 0.717 & 0.294 & 0.707 & 0.270 & 0.679 & 0.739 & 0.281\\
Systolic BP & -0.049 & 0.037 & -0.051 & 0.035 & -0.048 & 0.036 & -0.044 & 0.032 & -0.042 & -0.044 & 0.034\\
Pulse & 0.008 & 0.011 & 0.009 & 0.011 & 0.009 & 0.011 & 0.007 & 0.010 & 0.006 & 0.006 & 0.010\\
\midrule
\addlinespace[0.3em]
\multicolumn{12}{l}{\textbf{Predictions for Sam}}\\
\hspace{1em}PE & \multicolumn{2}{c}{0.032} & \multicolumn{2}{c}{0.036} & \multicolumn{2}{c}{0.036} & \multicolumn{2}{c}{0.044} & 0.050 &  & \\
\hspace{1em}PM (exact) & \multicolumn{2}{c}{0.036} & \multicolumn{2}{c}{0.040} & \multicolumn{2}{c}{0.040} & \multicolumn{2}{c}{0.048} &  & \multicolumn{2}{c}{0.042}\\
\hspace{1em}PM (MacKay) & \multicolumn{2}{c}{0.037} & \multicolumn{2}{c}{0.041} & \multicolumn{2}{c}{0.042} & \multicolumn{2}{c}{0.049} &  &  & \\
\hspace{1em}PM (self-proj.) & \multicolumn{2}{c}{0.036} & \multicolumn{2}{c}{0.040} & \multicolumn{2}{c}{0.041} & \multicolumn{2}{c}{0.048} &  &  & \\
\hspace{1em}95\% CrI & \multicolumn{2}{c}{(0.011, 0.085)} & \multicolumn{2}{c}{(0.013, 0.091)} & \multicolumn{2}{c}{(0.014, 0.090)} & \multicolumn{2}{c}{(0.019, 0.099)} &  & \multicolumn{2}{c}{(0.015, 0.086)}\\
\bottomrule
\end{tabular}
\begin{tablenotes}
\small
\item [] MI: Myocardial infarction; BP: Blood pressure; CV: Cross-validation; MLE: Maximum likelihood estimate;
\item [] PE: Plug-in estimate; PM: Posterior mean; SE: standard error; CrI: credible interval. PM (exact), PM (MacKay) and PM (self-proj.) are the posterior mean computed by Gauss-Hermite quadrature, MacKay's approximation, and self-projection, respectively. For the MCMC ridge, the exact PM is the posterior sample mean (and MacKay/self-projection, being Laplace-approximation computations, do not apply).
\item [] * Provided for comparison, as these are based on penalised MLEs under a frequentist framework, and therefore are not strictly Bayesian.
\end{tablenotes}
\end{threeparttable}
\end{table}

Several observations in this example are worth highlighting. First, the
effect of shrinkage priors is manifested as the coefficients of the
predictors being generally closer to zero compared with those of the
flat prior. The posterior means are further moved to the centre compared
to the plug-in predictions, reflecting the shrinkage properties of the
posterior mean. Overall, the choice of the prior and posterior mean
versus plug-in predictions had non-trivial effects for the exemplary
patient, with the largest predicted probability (0.050 - plug-in
prediction from the frequentist ridge) being 56\% higher than the
smallest (0.032 - plug-in prediction from the flat prior model), and
straddling the plausible 0.04 risk threshold.

\begin{figure}[!ht]
  \centering
  \caption{The difference between logit-transformed posterior mean (PM) associated with each prior and the linear term ($\eta$) as a function of $\eta$. Panel E shows the penalised MLE from the frequentist ridge, and panel F the posterior mean from the exact (MCMC) Bayesian ridge, for comparison.}
  \includegraphics[width=0.95\linewidth]{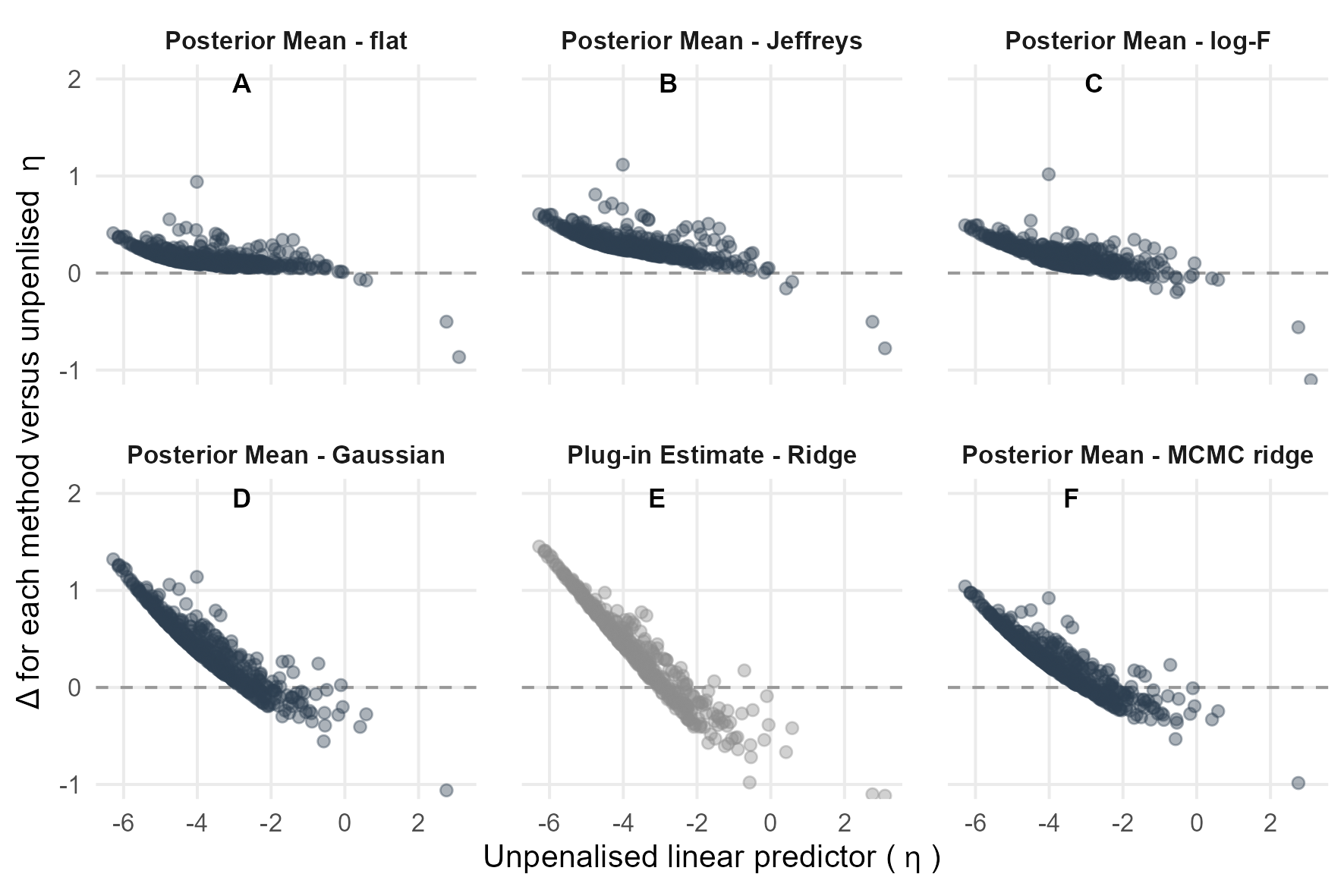}
  \caption*{\scriptsize The frequentist ridge (panel E) uses cross-validated tuning targeting deviance; the MCMC ridge (panel F) is the posterior mean from the fully Bayesian fit}
  \label{fig:shrinkage}
\end{figure}

Figure \ref{fig:shrinkage} showcases the shrinkage effect of predictions
in the sample. The X-axis is \(\eta\) in the development sample from the
flat-prior model (i.e., logit-transformed plug-in values), and the
Y-axis is the difference between logit-transformed posterior means and
\(\eta\)s. The shrinkage properties of the posterior mean result in a
positive difference when \(\eta\) is negative, and a negative difference
when it is positive (thus resulting in the narrower spread of \(\eta\)s
and the ensuing predicted probabilities). For comparison, panel E
presents the shrinkage effect from the frequentist ridge (with the
tuning parameter estimated via cross-validation, minimising deviance),
and panel F presents the posterior mean from the exact (MCMC) Bayesian
ridge.

\section{Discussion}\label{discussion}

This article has proposed and evaluated a pragmatic Bayesian approach
for developing and deploying prediction models that results in coherent
uncertainty propagation from the development sample to the point of
care, whilst also focusing on the benefit of using the posterior mean
for predictions. We placed a large emphasis on pragmatism, proposing an
approach that requires only minor modifications of commonly used
analytical tools, resulting in presentable, deployable, and
transportable models while remaining reproducible and computationally
tractable. This pragmatism rests on two main features: the use of priors
that enable Laplace approximation of the posterior, resulting in a model
being packaged as a vector of point estimates and a covariance matrix
for regression coefficients; and the proposal of algorithms for
deployment-time computation of posterior mean as the principled measure
of centrality. As part of the latter, we proposed a two-step regression
approach based on KL divergence minimisation that results in a simple
logistic equation approximating the posterior mean.

Our simulation studies generally confirmed the benefits of using such
shrinkage priors coupled with using the posterior mean, mainly in terms
of improvement in the clinical utility of predictions in small to medium
sample sizes for model development. In GUSTO-I simulations, this
approach significantly outperformed plug-in predictions from both
penalised and unpenalised likelihoods. Interestingly, they also
outperformed predictions from a MCMC-based Bayesian ridge model. This is
not a statement against exact Bayesian methods. Rather, it can be a
reminder that hierarchical shrinkage priors should not be seen as the
universal gold standard. As recommended, different classes of priors can
be examined in the initial steps of model development before finalising
the study design\citep{gelman_bayesian_2014}. If shrinkage priors that
enable Laplace approximation of the posterior perform well, the final
model can be represented as the parameters of multivariate normal
distribution for predictor effects - thus significantly facilitating
model presentation, deployment, and transportability. It can also
facilitate model aggregation across different studies, which can
substatially improve predictive performance\citep{debray_cpmma_2014}.

One could classify the first component of our proposal - the use of
shrinkage priors - as a form of likelihood penalisation, without
invoking Bayes. This characterisation, while technically correct, misses
three essential features of Bayesian thinking. First, whereas
uncertainty under the frequentist view refers to the properties of
repeated experiments, a Bayesian view provides direct probabilistic
statements about the unknown truth - the information the patient and
care provider seek. Second, while our focus was on single-population
learning, Bayesian thinking provides a solid bedrock for incorporating
external knowledge. For example, the posterior distribution from
previous development studies naturally becomes the prior for the new
analysis, ensuring optimal learning and coherent uncertainty
propagation. Finally, the expected utility framework provides a
principled justification for the posterior mean (a Bayesian quantity) as
the optimal metric for decision-making. It is not incidental that in our
simulations, posterior mean mostly outperformed the plug-in estimator in
terms of NB (with a degree of gain depending on where the threshold is
placed), as NB builds directly on an expected utility framework. Indeed,
in several simulated scenarios involving small-to-medium samples, such
changes gave rise to gains in clinical utility that plug-in predictions
could only match at noticeably larger sample sizes. Aligning the
deployed model predictions with decision theory can be an inexpensive
way of improving efficiency!

We presented the core developments around binary outcomes to keep this
paper focused, but the proposed approximate Bayesian approach is broadly
applicable to other outcome types. One can package the output of any GLM
as a vector of means, a covariance matrix, and a link function. Similar
principles apply to other modelling approaches such as survival models.
For proportional hazards (Cox) models, a technicality arises in that the
partial likelihood does not include the baseline hazard. However, one
may augment the parameter space to include the cumulative baseline
hazard at the time point of interest. Jeffreys prior is implementable
for any model with a computable Fisher information matrix; the log-F
prior can be implemented via data augmentation for categorical and
survival models\citep{greenland_logF_2015}; and the \(\texttt{mgcv}\)
implementation of the Bayesian ridge applies to all GLMs as well as to
proportional hazards
models\citep{wood_mgcvpackage_2000, wood_mgcv_2016}. In all these
approaches, Gauss-Hermite quadrature applies under the approximate
normality of the linear predictor. Our proposed two-step regression is
generic and can approximate the posterior mean from any of these
modelling strategies.

Several methodological extensions warrant further investigation. The
choice of priors and deployment-time computation methods for various
outcome types needs to be investigated. The Laplace approximation can be
further refined, via approaches such as the integrated nested Laplace
approximation (INLA)\citep{rue_inla_2009}. We explored INLA using
R-INLA\citep{lindgren_rinla_2017}, but this implementation reverted to
the ordinary Laplace approximation when a multivariate normal
approximation of the posterior was requested. Variational inference is
an information-theoretic alternative that approximates the posterior
with a parametric family by minimising KL
divergence\citep{blei_variational_2017}. This is conceptually attractive
as it is more aligned with the information-theoretic underpinnings of
statistical inference, though accessible implementations remain limited.
Model simplification and variable selection in a pragmatic Bayesian
framework also warrant investigation. In the frequentist paradigm,
shrinkage methods such as LASSO provide a data-driven means for variable
selection. Variable selection within a Bayesian framework can be
performed by projecting the reference model onto a smaller predictor
space\citep{mcLatchie_projpred_2025}. Projection predictive mapping has
primarily been developed in the context of MCMC-based approaches, and
its extension to the Laplace-approximation-based context needs to be
pursued. Summarising the posterior distribution as a multivariate
parametric distribution could in principle be applied to the output of
black-box models such as neural networks, provided their predictive
uncertainty can be represented through an appropriate parametric
distribution. Uncertainty quantification without loss of predictive
power for such models, however, might demand alternative approaches
beyond the scope of this work\citep{pmlr-v48-gal16}. Further, a Bayesian
view enables the extension of decision-theoretic approaches to
uncertainty assessment, referred to as value-of-information
analysis\citep{sadatsafavi_voipred_2022}, to the point of care - for
example to decide for which patients prediction uncertainty is large
enough to demand a more definitive test. Crucially, model
transportability can be a particularly promising area for Bayesian
approaches, with potential to improve the efficiency and coherence of
current practices.

We have summarised our specific recommendations in Table
\ref{tab:recommendations}. Our last recommendation might deserve further
reflection. Adopting a Bayesian framework requires rethinking what
counts as `good' performance. Frequentist estimation has long privileged
unbiasedness of predictions. An example is the O/E ratio, which is
guaranteed to be equal to 1 within the development sample when fitting a
logistic regression using (penalised or unpenalised) MLE. However, this
might come at the cost of overfitting and extreme predictions, higher
prediction error, worse information loss, and lower clinical utility. In
contrast, Bayesian decision theory constructs estimators targeting
explicit loss functions. It is a well-known result that with a proper
prior distribution, the posterior mean is generally biased in finite
samples\citetext{\citealp{bickel_bayes_1967}; \citealp[p244]{lehmann_theory_1998}; \citealp[p94]{gelman_bayesian_2014}};
yet it emerges as the optimal metric with respect to expected quadratic
prediction error, KL divergence, and, critically for risk prediction,
clinical utility. This line of reasoning indicates that under a Bayesian
paradigm, model assessment should prioritise efficient learning under
relevant objective functions, rather than the finite-sample bias of a
given summary statistic.

Clinical prediction models are, by their nature, about the transfer of
knowledge - from a development sample to the bedside, and from one
population to another. At the bedside, two important questions arise:
how uncertain evidence should be communicated, and how it should inform
medical decisions. The Bayesian approach provides coherent answers to
both. While adopting this view implies a paradigm shift in how we
conceptualise, present, and deploy prediction models, the key analytical
updates - specifying prior distributions and using the posterior mean
for decision making are achievable with minor tweaks to the current
machinery of risk prediction.

\begin{table}[ht]
\centering
\caption{Recommendations for a pragmatic Bayesian workflow.}
\renewcommand{\arraystretch}{1.2}

\begin{tabular*}{\textwidth}{@{\extracolsep{\fill}} p{3.5cm}  >{\normalsize}p{12cm}}
\toprule

\textbf{Recommendation} & \textbf{Justification} \\
\midrule

\textbf{Use a proper prior} & Unless the sample size is very large, we suggest some basic shrinkage priors to be used. This is both from our simulation studies showing the high variability of both plug-in (MLE) and posterior mean predictions under a flat prior, as well as the absurdity of an unpenalised model from a Bayesian view: that we are assuming all predictor effects are equally plausible. Priors like Jeffreys are widely implemented in regression routines, while a log-F prior can easily be incorporated via data augmentation.  \\

\addlinespace

\textbf{Use the posterior mean as a measure of centrality} & The posterior mean is the unique maximiser of expected utility; it enables straightforward and honest patient communication ("\textit{In people with the same characteristics as you, the expected risk is X\%}"). Deployment-time computations are not a barrier: if credible intervals are to be reported, the variance of the linear term is already calculated, making the additional cost of computing the posterior mean via quadrature or MacKay's approximation trivial. Self-projection further reduces posterior mean computation to a single inverse-link evaluation, similar to the plug-in prediction. \\

\addlinespace

\textbf{Quantify and communicate prediction uncertainty} & This is our general recommendation, much of which is aired previously\cite{riley_importance_2025}. A Bayesian framework, by characterising the full posterior distribution of predicted risk, allows communication of uncertainty in an intuitive way. Unlike frequentist interpretations which describe the outcomes of long-run hypothetical experiments, a Bayesian interpretation directly addresses the patient’s central questions such as "\textit{What is the probability that my personal risk lies within this range?}". \\

\addlinespace

\textbf{Consider the Bayesian alternative to bootstrapping penalised regression} & Bootstrapping CV-tuned ridge or LASSO is computationally expensive and introduces Monte Carlo noise. For LASSO, the zero-inflated empirical distribution of coefficients is awkward to summarise. The Bayesian ridge enables approximate Bayesian inference without these issues. Generally, combined with posterior mean's inherent shrinkage, milder shrinkage priors might provide comparable protection against overfitting while enabling proper uncertainty quantification. \\

\addlinespace

\textbf{Do not fixate on the O/E ratio} & The canonical-link MLE preserves O/E in the development sample by construction. As such, if the sample is representative of the population, MLE-based predictions are, on average, unbiased. The Bayesian approach, on the other hand, encourages defining summary predictions based on explicit objective functions (e.g., maximising clinical utility) and evaluating their performance accordingly - accepting that finite-sample unbiasedness may be sacrificed for better individual-level prediction.\\

\bottomrule
\end{tabular*}
\label{tab:recommendations}
\end{table}

\clearpage

\bibliography{references.bib}

\end{document}